\documentclass[a4paper,11pt]{article}
\usepackage{jheppub}
\makeatletter
\gdef\@fpheader{Accepted for publication in JHEP}
\makeatother
\usepackage[utf8]{inputenc}
\usepackage[section]{placeins}

\DeclareMathAlphabet{\mathbtt}{OT1}{lmtt}{b}{n}
\newcommand{\mi}[1]{\mathbtt{#1}}

\newcommand{\blk}{\mathfrak{b}}

\newcommand{\eq}[1]{\begin{align} #1 \end{align}}
\newcommand{\mean}[1]{\langle #1 \rangle}

\newcommand{\sNN}{\sqrt{s_{\rm NN}}}

\title{Subensemble Acceptance Method 3.0: General Corrections to Cumulants from Exact Conservation Constraints}

\author[a,b]{Roman Poberezhniuk,\note{Corresponding author.}}
\author[a,b]{Volodymyr A. Kuznietsov,}
\author[a]{Gr\'egoire Pihan}
\author[a]{and Volodymyr Vovchenko}

\affiliation[a]{Physics Department, University of Houston,\\Houston, TX 77204, USA}
\affiliation[b]{Bogolyubov Institute for Theoretical Physics,\\03680 Kyiv, Ukraine}

\emailAdd{rpoberez@central.uh.edu}

\abstract{We present the subensemble acceptance method 3.0 (SAM-3.0), which corrects cumulants of an observable measured in a subsystem of a large system for the effect of exact global conservation of multiple charges.
The required input is the set of joint grand-canonical cumulants of the acceptance observable with the total event charges, from which the canonical cumulants follow algebraically via a closed recursion based on (multivariate) partial exponential Bell polynomials.
The framework accommodates any number of observables, including non-conserved quantities such as net protons, and any number of simultaneously conserved charges, including the total energy, which yields the microcanonical ensemble.
The mapping contains SAM-1.0 and SAM-2.0 as special cases and, unlike SAM-2.0, reproduces the exact binomial-acceptance limit.
We also derive the leading finite-size corrections from the saddle-point expansion.
We apply the method to update the hydrodynamics-based non-critical baseline (Hydro-EV) for net-proton cumulants at RHIC-BES energies, finding a refined baseline that agrees with direct canonical Monte Carlo sampling and stays close to the earlier SAM-2.0 result. We further validate the formalism against direct Monte Carlo sampling with exact simultaneous conservation of baryon number, electric charge, and strangeness, including hadronic-afterburner effects.}


\begin{document}
\maketitle
\flushbottom

\section{Introduction}

Event-by-event fluctuations of conserved charges are key observables to probe the QCD phase structure in heavy-ion collisions.
Higher-order cumulants of net-baryon, net-charge, and net-strangeness distributions are expected to be sensitive to the QCD phase structure at finite density, including the possible QCD critical point~\cite{Stephanov:1998dy,Stephanov:1999zu,Stephanov:2008qz,Stephanov:2011pb}, and provide a potential direct link to first-principles lattice QCD calculations, where they are accessible through the corresponding susceptibilities~\cite{Bazavov:2017dus,Borsanyi:2018grb}.
This has motivated extensive experimental efforts at RHIC, the LHC, and other facilities.
The STAR Collaboration has measured net-proton cumulants across a broad range of collision energies in the Beam Energy Scan (BES) program~\cite{Adamczyk:2013dal,Adam:2020unf,STAR:2021iop}, with the recent BES-II results~\cite{STAR:2025zdq} providing substantially improved precision.
Complementary measurements have been performed by HADES at lower energies~\cite{HADES:2020wpc} and by ALICE at the LHC~\cite{ALICE:2019nbs,ALICE:2022xpf}.

On the theory side, predictions for fluctuation observables are predominantly formulated in terms of the grand-canonical ensemble (GCE).
This is the natural framework for lattice QCD calculations of susceptibilities~\cite{Bazavov:2017dus,Borsanyi:2018grb}, for effective theory predictions near the QCD critical point~\cite{Stephanov:2008qz,Stephanov:2011pb}, 
and calculations within functional methods~\cite{Isserstedt:2019pgx,Fu:2019hdw,Gao:2020qsj}.
Projection of energy and conserved-charge susceptibilities onto measurable hadron numbers was recently developed using a maximum-entropy approach to chemical freeze-out of fluctuations~\cite{Pradeep:2022eil}, with early applications restricted to equilibrium GCE calculations~\cite{Karthein:2025hvl}.
Likewise, Monte Carlo event generators and hydrodynamic simulations typically operate in the grand-canonical ensemble: Cooper-Frye particlization samples hadrons from a grand-canonical thermal distribution, and while samplers that incorporate global conservation do exist~\cite{Schwarz:2017bdg,Oliinychenko:2019zfk,Vovchenko:2020kwg}, they add considerable complexity and computational cost.

In a real collision, however, conserved charges such as net baryon number $B$, electric charge $Q$, and strangeness $S$ are exactly conserved globally.
On the experimental side, detectors measure only a subsystem (acceptance) of the full final state, and while this may approximate a grand-canonical setup~\cite{Koch:2008ia}, the global conservation law remains in force.
Therefore, one needs a practical framework to relate grand-canonical cumulants -- the natural output of theory -- to those measured in a subsystem subject to exact global conservation.

The role of exact conservation laws as an essential baseline for fluctuation observables has been recognized in several works.
Early studies investigated conservation effects on particle number fluctuations in non-interacting systems, including the HRG model in the canonical ensemble for second-order correlations and fluctuations~\cite{Begun:2004gs}, the analysis of baryon-number conservation effects on high-order net-proton cumulants~\cite{Bzdak:2012an,Braun-Munzinger:2016yjz}, and the calculation of an HRG canonical ensemble (HRG CE) baseline for net-proton cumulant measurements~\cite{Braun-Munzinger:2020jbk}.
Recently, correlations induced by exact conservation of energy and momentum, alongside the conserved charges, were studied within hydrodynamics~\cite{Savchuk:2023yeh} and for momentum modes of an ideal gas~\cite{Jaiswal:2026cxt}.

The subensemble acceptance method (SAM) was developed to go beyond the ideal gas and incorporate correlations arising from any equation of state.
SAM-1.0~\cite{Vovchenko:2020tsr} relates cumulants inside a given coordinate-space acceptance to grand-canonical susceptibilities in a spatially uniform thermal system with a globally conserved charge, under the assumption of short-range correlations.
SAM-1.0 was tested within a mean-field (van der Waals) model of critical-point fluctuations, where the combined effects of finite system size and global charge conservation were analyzed~\cite{Poberezhnyuk:2020ayn}.
In Ref.~\cite{Vovchenko:2020gne}, the method was extended to multiple conserved charges.
An alternative derivation was presented in Ref.~\cite{Barej:2022jij} and next-to-leading-order (NLO) corrections with respect to system size were worked out in~\cite{Barej:2022ccb}.
SAM-2.0~\cite{Vovchenko:2021yen} relaxed several of these assumptions: it treated non-uniform systems, momentum-space acceptances, and non-conserved quantities (such as net-proton number).
SAM-2.0 underpinned the hydrodynamics-based calculation of the excluded-volume (Hydro-EV) baseline for proton number cumulants at RHIC-BES~\cite{Vovchenko:2021kxx} and LHC~\cite{Vovchenko:2020kwg}.

SAM-2.0 incorporates the canonical constraint $\delta_{B_{\rm tot},B_0}$
into the grand-canonical distribution.
Here $B_{\rm tot} = B_{\rm in} + B_{\rm out}$ is the total conserved charge of the system, with $B_{\rm in}$ and $B_{\rm out}$ denoting its parts contained in the acceptance and in its complement, respectively, and $B_0$ is the fixed value of the total charge.
The method relies on
two key assumptions:
(i) $B_{\rm in}$ and $B_{\rm out}$ are independent in the unconstrained reference ensemble,
$P_{\rm gce}(B_{\rm in},B_{\rm out})
 = P^{\rm gce}_{\rm in}(B_{\rm in})\, P^{\rm gce}_{\rm out}(B_{\rm out})$;
(ii) the distributions $P^{\rm gce}_{\rm in/out}$ are unimodal and sharply
peaked, so that a saddle-point approximation can be applied.
A drawback of SAM-2.0 is that the in/out independence assumption~(i), while valid for coordinate-space acceptances when particle correlations are short-range, may be questionable for momentum-space acceptances.
Ref.~\cite{Vovchenko:2021yen} illustrated this limitation by considering a binomial momentum-space acceptance in a static thermal system:
the exact canonical cumulants of the accepted charge reduce to those of a binomial distribution,
whereas SAM-2.0 incorrectly predicts non-trivial (non-binomial) behavior, an artifact of the independence assumption.

In this work we address this limitation by reformulating the SAM framework in terms of the joint cumulant generating function of the observable~$X$ and the total conserved charge~$B_{\rm tot}$, rather than treating the ``in'' and ``out'' subsystems separately.
The conditional integral is evaluated through the saddle-point approximation, as in SAM-2.0, but without the in/out factorization assumption.
Arbitrary correlations between $B_{\rm in}$ and $B_{\rm out}$ in the reference ensemble are then encoded in the mixed grand-canonical cumulants $\kappa^{XB}_{n\,m}$.
In particular, the intermediate layer of shape functions and their derivatives used in SAM-2.0 is eliminated entirely, and the canonical cumulants are expressed directly in terms of the grand-canonical mixed cumulants $\kappa^{XB}_{n\,m}$.
The resulting SAM-3.0 mapping satisfies the binomial acceptance benchmark exactly and reduces to SAM-2.0 when in/out correlations vanish.
We extend the framework to multiple observables and multiple exactly conserved charges using multivariate Fa\`a di Bruno coefficients (partial exponential Bell polynomials).
We additionally derive next-to-leading-order finite-size corrections to the canonical cumulants from the saddle-point expansion.
The exact conservation of the total energy can be incorporated by treating the energy as an additional conserved charge, yielding the microcanonical ensemble (MCE) correction.

The paper is organized as follows.
Section~\ref{sec:formalism} introduces the SAM-3.0 formalism and derives the expressions for cumulants of a single observable correlated to a single conserved charge.
Section~\ref{sec:multidim} extends the framework to multiple observables and conserved charges.
Section~\ref{sec:applications} discusses special cases and benchmarks, including the SAM-1.0 and SAM-2.0 limits, the binomial acceptance benchmark,
and the case of the microcanonical ensemble with exact conservation of total energy in addition to conserved charges.
Section~\ref{sec:numerical} presents numerical applications: a refined non-critical baseline (Hydro-EV~\cite{Vovchenko:2021kxx}) for net-proton cumulants at RHIC-BES using SAM-3.0, and a validation of the formalism against direct canonical Monte Carlo sampling with exact conservation of conserved charges.
Summary and outlook are presented in Sec.~\ref{sec:summary}.
Appendix~\ref{sec:finite_size} derives the next-to-leading-order finite-size corrections, 
Appendix~\ref{sec:nlo_numerical} quantifies the magnitude of NLO corrections in the Hydro-EV setup, and Appendix~\ref{sec:binomial_multi} derives the general multidimensional binomial acceptance framework with species-dependent acceptance probabilities and arbitrary observables used in Sec.~\ref{sec:binomial_limit}.

\section{Formalism}
\label{sec:formalism}

\subsection{Setup}

Let $X$ denote the random variable of interest (for instance,
net-proton number in a given momentum-space acceptance) and let $B_{\rm tot}( = B_{\rm in} + B_{\rm out})$
denote the exactly conserved total charge to which $X$ is correlated (for
instance, net baryon number of the full event).  Denote by
\begin{equation}
  P_{\rm gce}(X,B_{\rm tot})
\end{equation}
the joint distribution in some unconstrained reference ensemble, which we call
``grand-canonical'' for brevity, and by
\begin{equation}
  P_{\rm gce}(B_{\rm tot}) = \sum_X P_{\rm gce}(X,B_{\rm tot})
\end{equation}
the corresponding marginal of the total charge.
The associated cumulant-generating function\footnote{We use exponential-generating-function conventions throughout, i.e.\ Taylor expansions are written with factorials.} reads
\eq{
G_{\rm gce}(t_X, t_B) &= \ln \mean{e^{t_X X + t_B B_{\rm tot}}}_{\rm gce} = \ln \sum_{X,B_{\rm tot}} P_{\rm gce}(X,B_{\rm tot}) e^{t_X X + t_B B_{\rm tot}} \nonumber\\
&= \sum_{n,m} \frac{\kappa^{XB}_{n\,m}}{n! m!} (t_X)^n (t_B)^m
}
where $\kappa^{XB}_{n\,m}$ denotes the joint GCE cumulant of order $n$ in the observable~$X$ and order $m$ in the conserved charge~$B_{\rm tot}$; we write $\kappa^{X}_{n}\equiv\kappa^{XB}_{n\,0}$ and $\kappa^{B}_{m}\equiv\kappa^{XB}_{0\,m}$ for the marginal cumulants.
Throughout the paper we use the following notation conventions for cumulants: an unmarked $\kappa$ denotes a grand-canonical cumulant; the upper label lists each random variable exactly once and the subscript gives the order in each, so that $\kappa^{XB}_{n\,m}$, $\kappa^{XYBQ}_{n_1n_2m_1m_2}$, etc. are joint GCE cumulants. Canonical (constrained) cumulants at fixed total charge are written with a conditional bar, $\kappa^{X|B}_n$, $\kappa^{XY|BQ}_{nm}$.
The $\mean{\cdot}_{\rm gce}$ denotes GCE averaging: $\mean{f(X,B_{\rm tot})}_{\rm gce}\equiv\sum_{X,B_{\rm tot}}P_{\rm gce}(X,B_{\rm tot})\,f(X,B_{\rm tot})$.

Then the probability
distribution of $X$ in the presence of exact global conservation of
$B_{\rm tot} = B_0$ is simply the conditional probability
\begin{equation}
  \label{eq:Pce_cond}
  P_{\rm ce}(X)
  \equiv P_{\rm gce}\bigl(X \,\big|\, B_{\rm tot}=B_0\bigr)
  = \frac{P_{\rm gce}(X,B_{\rm tot}=B_0)}
         {P_{\rm gce}(B_{\rm tot}=B_0)}.
\end{equation}
In other words, the canonical ensemble is the grand-canonical ensemble
conditioned on a fixed value of the total conserved charge.
In the context of a thermal system, this is true for any Abelian conserved charge, such as baryon number, electric charge, or strangeness in QCD.

The corresponding canonical cumulant generating function (CGF) reads
\begin{equation}
  \label{eq:G_ce}
  G_{\rm ce}(t)
  = \ln \mean{e^{t X}}_{\rm ce}
  = \ln \frac{\mean{e^{t X}\,\delta_{B_{\rm tot},B_0}}_{\rm gce}}
         {\mean{\delta_{B_{\rm tot},B_0}}_{\rm gce}}
  = \sum_{n} \frac{\kappa^{X|B}_n}{n!} t^n,
\end{equation}
where the second equality follows from the conditional-probability definition~\eqref{eq:Pce_cond} of $P_{\rm ce}(X)$ with the Kronecker delta $\delta_{B_{\rm tot},B_0}$ enforcing $B_{\rm tot}=B_0$, and the last is the definition of the canonical cumulants $\kappa^{X|B}_n$.

All canonical cumulants $\kappa^{X|B}_n$ of $X$ at fixed global charge $B_{\rm tot}=B_0$ are
then obtained as derivatives of $G_{\rm ce}(t)$ at $t=0$.
The central question is how to express these canonical cumulants in terms of
the \emph{grand-canonical} information (cumulants $\kappa^{XB}_{n\,m}$) about the system.

SAM-2.0~\cite{Vovchenko:2021yen} provides one such mapping under the assumption that $B_{\rm in}$ and $B_{\rm out}$ (the accepted and rejected portions of the conserved charge, $B_{\rm tot}=B_{\rm in}+B_{\rm out}$) are independent in the reference ensemble.
Here we reformulate the problem (SAM-3.0) by working directly with the joint CGF $G_{\rm gce}(t_X,t_B)$ and imposing conservation at the level of this two-variable generating function.
As discussed in the Introduction, this captures arbitrary in/out correlations in the reference ensemble through the joint cumulants $\kappa^{XB}_{n\,m}$, eliminating the shape-function construction of SAM-2.0.

Using the Fourier representation of the Kronecker delta,
\begin{equation}
  \delta_{n,m}
  = \int_0^{2\pi} \frac{d\phi}{2\pi}\,e^{i\phi(n-m)},
\end{equation}
we obtain
\begin{align}
  \mean{e^{t X}\,\delta_{B_{\rm tot},B_0}}_{\rm gce}
  &= \int_0^{2\pi} \frac{d\phi}{2\pi}\,
     e^{i\phi B_0}\,
     \mean{e^{t X - i\phi B_{\rm tot}}}_{\rm gce}
  = \int_0^{2\pi} \frac{d\phi}{2\pi}\,
     \exp\bigl[i\phi B_0 + G_{\rm gce}(t,-i\phi)\bigr],\\[1ex]
  \mean{\delta_{B_{\rm tot},B_0}}_{\rm gce}
  &= \int_0^{2\pi} \frac{d\phi}{2\pi}\,
     e^{i\phi B_0}\,
     \mean{e^{- i\phi B_{\rm tot}}}_{\rm gce}
  = \int_0^{2\pi} \frac{d\phi}{2\pi}\,
     \exp\bigl[i\phi B_0 + G_{\rm gce}(0,-i\phi)\bigr] .
\end{align}
Thus the canonical CGF has the exact integral representation
\begin{equation}
  \label{eq:Gce_integral}
  G_{\rm ce}(t)
  = \ln\frac{
      \displaystyle
      \int_0^{2\pi}\frac{d\phi}{2\pi}\,
      \exp\bigl[i\phi B_0 + G_{\rm gce}(t,-i\phi)\bigr]
    }{
      \displaystyle
      \int_0^{2\pi}\frac{d\phi}{2\pi}\,
      \exp\bigl[i\phi B_0 + G_{\rm gce}(0,-i\phi)\bigr]
    }.
\end{equation}

\subsection{Saddle-point approximation}
\label{sec:saddlepoint}

The integrals in Eq.~\eqref{eq:Gce_integral} are amenable to a saddle-point (steepest-descent) approximation when the distribution $P_{\rm gce}(B_{\rm tot})$ is sharply peaked and unimodal, as in the single-phase thermodynamic limit or in high-multiplicity events from Cooper-Frye particlization on a hydrodynamic hypersurface or subsequent hadronic afterburners.

Introduce a complex variable
\begin{equation}
  \lambda = -i\phi
\end{equation}
and consider a contour deformation such that the integrals over $\phi$ are expressed as
\begin{equation}
\label{eq:contour}
  I(t) \equiv \int_{\cal C}
  \frac{d\lambda}{2\pi i}\,
  \exp\bigl[F(t,\lambda)\bigr],
  \qquad
  F(t,\lambda) = G_{\rm gce}(t,\lambda) - \lambda B_0.
\end{equation}
The canonical CGF then reads
\begin{equation}
  G_{\rm ce}(t)
  = \ln I(t) - \ln I(0).
\end{equation}
The integrals are dominated by a saddle point
$\lambda_\star(t)$ defined by
\begin{equation}
\label{eq:saddle}
  \left.\frac{\partial F}{\partial\lambda}\right|_{\lambda=\lambda_\star(t)}
  = 0
  \quad\Rightarrow\quad
  \frac{\partial G_{\rm gce}}{\partial\lambda}(t,\lambda_\star(t)) = B_0.
\end{equation}

The contour deformation is justified by the properties of $Z(t,\lambda)\equiv\mean{e^{tX+\lambda B_{\rm tot}}}_{\rm gce}$, to which the integrand is proportional, $e^{F(t,\lambda)}=e^{-\lambda B_0}\,Z(t,\lambda)$. 
Provided $Z(t,\lambda)$ is finite for real $\lambda$ in an open interval containing 0 and $\lambda_\star(t)$, it is analytic in the corresponding vertical strip.
For integer-valued \(B_{\rm tot}\) (and hence \(B_0\)), the full integrand \(e^{-\lambda B_0}Z(t,\lambda)\) is \(2\pi i\)-periodic.
The original contour, oriented from $-2\pi i$ to $0$ in Eq.~\eqref{eq:contour}, can therefore be translated by Cauchy's theorem to pass through any real $\lambda$ within the strip, the horizontal edge contributions canceling by periodicity. For real \(t\), strict convexity of $G_{\rm gce}$ in real $\lambda$, which follows from $\partial^2_\lambda G_{\rm gce}$ being the variance of $B_{\rm tot}$ in the reweighted ensemble, implies that there can be at most one real saddle. In our construction 
$\lambda_\star(0) = 0$, and a unique nearby saddle exists for sufficiently small $t$ provided $\kappa_2^B > 0$.
The validity of the Gaussian saddle-point expansion is assessed through the standardized higher cumulants introduced in the following paragraph.

We introduce $\delta\lambda\equiv \lambda-\lambda_\star(t)$ and expand
\begin{equation}
  \label{eq:F_Taylor}
  F\!\left(t,\lambda_\star+\delta\lambda\right)
  =
  F_0(t)
  +\sum_{n\ge2}\frac{1}{n!}\,F_n(t)\,(\delta\lambda)^n,
\end{equation}
where $F_0(t)\equiv F\!\left(t,\lambda_\star(t)\right)$, $F_n(t)\equiv \partial_\lambda^{\,n}G_{\rm gce}(t,\lambda)\big|_{\lambda=\lambda_\star(t)}$ for $n\ge 2$ (since $-\lambda B_0$ is linear in $\lambda$), and the linear term vanishes by the saddle condition~(\ref{eq:saddle}).
The quadratic term of the expansion~\eqref{eq:F_Taylor} limits the integration to a region of width $|\delta\lambda|\sim(\kappa^B_2)^{-1/2}$ around the saddle, within which the term of order $n$ contributes at relative order $\kappa^B_n/(\kappa^B_2)^{n/2}$.
The saddle-point approximation, which retains the quadratic term alone, is therefore valid when the standardized cumulants of the conserved charge are small, $\kappa^B_n/(\kappa^B_2)^{n/2}\to 0$ for $n\ge 3$.
In an extensive system $\kappa^B_n\propto V$ and the condition reduces to $\kappa^B_2\gg 1$.
Note that the condition constrains only the ratios $\kappa^B_n/(\kappa^B_2)^{n/2}$ rather than the cumulants $\kappa^B_n$ themselves. 
When $\kappa^B_n = O(V)$ for all fixed $n$, as in a conventional single-phase extensive system away from phase coexistence, these ratios scale as $\mathcal{O}\big[(\kappa^B_2)^{1-n/2}\big]$ and the saddle expansion becomes systematically controlled at large $\kappa_2^B$, with the finite-size corrections organized in powers of $1/\kappa_2^B$.
The Gaussian saddle-point approximation can become unreliable in small systems with \(\kappa_2^B=O(1)\) or near a first-order phase transition, where \(P_{\rm gce}(B_{\rm tot})\) can become bimodal. In such regimes, the standardized higher cumulants \(\kappa_n^B/(\kappa_2^B)^{n/2}\), \(n\geq3\), need not become small. When these ratios are not small, the Gaussian truncation of the saddle-point expansion is not controlled.

Neglecting next-to-leading $\mathcal{O}(\ln V)$ corrections from the Gaussian
prefactor of the saddle-point integration,
we approximate the integral by its saddle value,
\(\ln I(t)\simeq F_0(t)\) (up to a \(t\)-independent constant), so that
\(G_{\rm ce}(t)\simeq F_0(t)-F_0(0)\).
Keeping the Gaussian prefactor yields the leading finite-size correction (see Appendix~\ref{sec:finite_size}).

The canonical CGF~\eqref{eq:Gce_integral} depends only on the reference GCE cumulant generating function $G_{\rm gce}(t,\lambda)$ and the target charge $B_0$, not on where we center the variable $\lambda$: shifting $\lambda\to\lambda+c$ moves the saddle $\lambda_\star(t)\to\lambda_\star(t)-c$ and leaves $G_{\rm ce}(t)$ invariant.
We use this freedom to place the $t=0$ saddle at the origin, $\lambda_\star(0)=0$, which by Eq.~\eqref{eq:saddle} requires the reference ensemble to already carry the target charge on average, $\langle B_{\rm tot}\rangle_{\rm gce}=\kappa^B_1=B_0$.\footnote{If the reference ensemble does not carry the target charge on average, $\langle B_{\rm tot}\rangle_{\rm gce}\neq B_0$, the $t=0$ saddle sits at some $\lambda_0\neq 0$ fixed by $\partial_\lambda G_{\rm gce}(0,\lambda_0)=B_0$, and the expansion below proceeds around $\lambda_0$ instead of the origin, leaving the canonical cumulants unchanged.}
With this choice, $\lambda_\star(0)=0$, so that $F(0,0)=0$ and $G_{\rm ce}(t)$ simplifies to
\begin{equation}
  \label{eq:Gce_final}
  G_{\rm ce}(t)
  = G_{\rm gce}[t,\lambda(t)] - \lambda(t) B_0,
\end{equation}
where we have written $\lambda(t)\equiv \lambda_\star(t)$ for
brevity.  The function $\lambda(t)$ is determined implicitly by the
saddle-point condition
\begin{equation}
  \label{eq:saddle_eq}
  \frac{\partial G_{\rm gce}}{\partial\lambda}(t,\lambda(t)) = B_0,
  \qquad
  \lambda(0) = 0.
\end{equation}

The canonical cumulants of $X$ follow from Eq.~\eqref{eq:Gce_final} as
\begin{equation}
\label{eq:dGdt}
  \kappa^{X|B}_n
  = \left.\frac{d^n G_{\rm ce}}{dt^n}\right|_{t=0}.
\end{equation}
To compute them, one needs the derivatives of the implicit solution
$\lambda(t)$. 

Below, we outline an iterative procedure to compute the derivatives of $\lambda(t)$ and the cumulants $\kappa^{X|B}_n$ based
on Bell polynomials.

\subsection{Expansion of \texorpdfstring{$\lambda(t)$}{lambda(t)}}

We expand the saddle-point solution $\lambda(t)$ in a formal power series in $t$:
\begin{equation}
  \label{eq:lambda_series}
  \lambda(t)
  = \sum_{n = 1}^\infty \frac{\lambda_n}{n!}\,t^n,
  \qquad
  \lambda_n \equiv \left.\frac{d^n\lambda}{dt ^n}\right|_{t=0}.
\end{equation}

The saddle-point condition reads
\eq{\label{eq:Sonedim}
S(t) \equiv \partial_\lambda G_{\rm gce}(t,\lambda(t)) - B_0 = 0.
}
Differentiating the cumulant generating function $G_{\rm gce}$ in $\lambda$ gives
\eq{
\partial_\lambda G_{\rm gce}(t,\lambda)
  = \sum_{n,m\ge 0}
    \frac{\kappa^{XB}_{n,m+1}}{n!\,m!}\,t^n\lambda^m.
}
This condition implies that all derivatives of $S(t)$ vanish at $t=0$:
\eq{
\frac{d^n S(t)}{dt^n}\bigg|_{t=0} = 0 \quad \forall n \ge 1.
}

To evaluate these derivatives we use the following identity for derivatives of a composite argument 
(Fa\`a di Bruno formula; see Ref.~\cite{leipnik2007multivariate}, Eq.~$(3.5)$, for a general case of the equation below)
\eq{\label{eq:FdB}
\frac{d^n}{dt^n} F[t,b(t)]
= F_{n,0} + \sum_{m=0}^n \sum_{q = 1}^{n-m} \binom{n}{m} F_{m,q} B_{n-m,q}(b_1,\ldots,b_{n-m-q+1}).
}
{\sloppy Here $F_{m,q}$ are the partial derivatives of $F(t)$ with respect to $t$ and $b(t)$, $B_{n-m,q}(b_1,\ldots,b_{n-m-q+1})$ are the partial exponential Bell polynomials, and $b_n = d^n b(t) / d t^n$.\par}

The $n$th order derivative of $S(t)$ is then
\eq{
\frac{d^n S(t)}{dt^n}
= \kappa^{XB}_{n\,1} + \sum_{m=0}^n \sum_{q = 1}^{n-m} \binom{n}{m} \kappa^{XB}_{m,q+1} B_{n-m,q}(\lambda_1,\ldots,\lambda_{n-m-q+1}).
}

Within the sum above, $\lambda_n$ can appear only through the Bell polynomial's highest-indexed argument: since $B_{n-m,q}$ depends on $\lambda_1,\dots,\lambda_{n-m-q+1}$, the constraint $n-m-q+1=n$ forces $(m,q)=(0,1)$, for which $B_{n,1}(\lambda_1,\dots,\lambda_n)=\lambda_n$ and $\kappa^{XB}_{0\,2}=\kappa^{B}_{2}$. Defining the remainder
\eq{\label{eq:Rn_def}
R_n \equiv \sum_{m=0}^n \sum_{\substack{q = 1\\(m,q)\neq(0,1)}}^{n-m} \binom{n}{m} \kappa^{XB}_{m,q+1} B_{n-m,q}(\lambda_1,\ldots,\lambda_{n-m-q+1}),
}
which involves only $\lambda_k$ with $k<n$, we can write
\eq{
\frac{d^n S(t)}{dt^n}
= \kappa^{XB}_{n\,1} + \lambda_n \kappa^{B}_{2} + R_n.
}
Since $d^n S(t)/dt^n = 0$ for $n\ge 1$, we solve iteratively:
\eq{\label{eq:lambda_n}
\lambda_n = -\frac{\kappa^{XB}_{n\,1} + R_n}{\kappa^{B}_{2}},
}
with the recursion proceeding in increasing $n$.

For illustration, we list the first few coefficients explicitly: 
\eq{\label{eq:lambda-result-1}
  \lambda_1 & = -\frac{\kappa^{XB}_{1\,1}}{\kappa^{B}_{2}}, \\
  \lambda_2 & = \frac{2 \kappa^{B}_{2} \kappa^{XB}_{1\,1} \kappa^{XB}_{1\,2} - \bigl(\kappa^{B}_{2}\bigr)^2 \kappa^{XB}_{2\,1} - \kappa^{B}_{3} \bigl(\kappa^{XB}_{1\,1}\bigr)^2}{\bigl(\kappa^{B}_{2}\bigr)^3}, \\\nonumber
  \lambda_3 & =\frac{1}{\bigl(\kappa^{B}_{2}\bigr)^4}\Bigl[ \kappa^{B}_{4}\bigl(\kappa^{XB}_{1\,1}\bigr)^3 -\kappa^{B}_{2}\Bigl( \kappa^{XB}_{1\,1}\bigl(6(\kappa^{XB}_{1\,2})^2 -3\kappa^{B}_{2}\kappa^{XB}_{2\,2}\bigr)  +\kappa^{B}_{2}\bigl(\kappa^{B}_{2}\kappa^{XB}_{3\,1} -3\kappa^{XB}_{1\,2}\kappa^{XB}_{2\,1}\bigr) \\ & +3\bigl(\kappa^{XB}_{1\,1}\bigr)^2\kappa^{XB}_{1\,3} \Bigr) -3\kappa^{B}_{3}\kappa^{XB}_{1\,1} \bigl(\kappa^{B}_{2}\kappa^{XB}_{2\,1} -3\kappa^{XB}_{1\,1}\kappa^{XB}_{1\,2}\bigr) \Bigr] \;-\;\frac{3\bigl(\kappa^{B}_{3}\bigr)^2\bigl(\kappa^{XB}_{1\,1}\bigr)^3} {\bigl(\kappa^{B}_{2}\bigr)^5}\, .\label{eq:lambda-result-3}
}

\subsection{Iterative solution for \texorpdfstring{$\kappa^{X|B}_n$}{the canonical cumulants}}

The constrained cumulants $\kappa^{X|B}_n$ follow from computing derivatives of the canonical CGF in Eq.~\eqref{eq:Gce_final}. Using the identity \eqref{eq:FdB}, one finds that the $(m,q)=(0,1)$ term in the sum yields $\kappa^{B}_{1}\lambda_n = B_0\lambda_n$, which cancels the $-\lambda_n B_0$ contribution from $-\lambda(t)B_0$, giving
\eq{\label{eq:kappa-ce-general}
\kappa^{X|B}_n
= \kappa^{X}_{n} + \sum_{m=0}^n \sum_{\substack{q = 1\\(m,q)\neq(0,1)}}^{n-m} \binom{n}{m} \kappa^{XB}_{m\,q}B_{n-m,q}(\lambda_1,\ldots,\lambda_{n-m-q+1}).
}

Explicit expressions for the first few cumulants read:
\begin{align}\label{eq:k1ce}
  \kappa^{X|B}_1 &= \kappa^{X}_{1},\\[1ex]
  \kappa^{X|B}_2
  &= \kappa^{X}_{2} - \frac{(\kappa^{XB}_{1\,1})^2}{\kappa^{B}_{2}}\label{eq:k2ce},
  \\[1ex]
  \kappa^{X|B}_3 
  &= \kappa^{X}_{3}
   - 3\,\frac{\kappa^{XB}_{1\,1} \kappa^{XB}_{2\,1}}{\kappa^{B}_{2}}
   + 3\,\frac{(\kappa^{XB}_{1\,1})^2 \kappa^{XB}_{1\,2}}{(\kappa^{B}_{2})^2}
   - \frac{(\kappa^{XB}_{1\,1})^3 \kappa^{B}_{3}}{(\kappa^{B}_{2})^3}, 
\\[1ex]
\kappa^{X|B}_4
&= \kappa^{X}_{4}
+ \frac{1}{\bigl(\kappa^{B}_{2}\bigr)^4}\Biggl\{
\bigl(\kappa^{XB}_{1\,1}\bigr)^2
\Bigl[
-6\,\kappa^{B}_{3}\Bigl(\kappa^{B}_{2}\kappa^{XB}_{2\,1}-2\,\kappa^{XB}_{1\,1}\kappa^{XB}_{1\,2}\Bigr)
+ \kappa^{B}_{4}\bigl(\kappa^{XB}_{1\,1}\bigr)^2
\Bigr]
\nonumber\\
&\qquad
-\kappa^{B}_{2}\Bigl[
4\,\kappa^{B}_{2}\kappa^{XB}_{1\,1}
\Bigl(\kappa^{B}_{2}\kappa^{XB}_{3\,1}-3\,\kappa^{XB}_{1\,2}\kappa^{XB}_{2\,1}\Bigr)
+6\,\bigl(\kappa^{XB}_{1\,1}\bigr)^2
\Bigl(2\bigl(\kappa^{XB}_{1\,2}\bigr)^2-\kappa^{B}_{2}\kappa^{XB}_{2\,2}\Bigr)
\nonumber\\
&\qquad\qquad\qquad
+3\,\bigl(\kappa^{B}_{2}\bigr)^2\bigl(\kappa^{XB}_{2\,1}\bigr)^2
+4\,\bigl(\kappa^{XB}_{1\,1}\bigr)^3\kappa^{XB}_{1\,3}
\Bigr]\Biggr\}
-3\,\frac{\bigl(\kappa^{B}_{3}\bigr)^2\bigl(\kappa^{XB}_{1\,1}\bigr)^4}{\bigl(\kappa^{B}_{2}\bigr)^5}.\label{eq:k4ce}
\end{align}

These formulas provide a direct mapping from GCE mixed cumulants of $(X,B_{\rm tot})$
to canonical cumulants of $X$ at fixed $B_{\rm tot} = B_0$, within the saddle-point
approximation and given the joint GCE cumulants up to the desired order.
Higher-order expressions can be generated using the accompanying \textsc{Mathematica} notebook~\cite{SAM3code}.

\section{Multiple observables and conserved charges}
\label{sec:multidim}

We now generalize the one-dimensional construction of Sec.~\ref{sec:formalism} to the case of
$d$ (possibly correlated) observables and $s$ exactly conserved charges.
We denote by ${\bf X}=(X_{1},\dots,X_{d})$ the vector of observables in the acceptance,
and by ${\bf B}_{\mathrm{tot}}=(B^{1}_{\rm tot},\dots,B^{s}_{\rm tot})$ the vector of globally conserved charges of the full event,
conditioned on fixed values ${\bf B}_{\mathrm{tot}}={\bf B}_{0}$.
The source vectors are ${\bf t}=(t_{1},\dots,t_{d})$, conjugate to ${\bf X}$, and $\boldsymbol{\lambda}=(\lambda_{1},\dots,\lambda_{s})$, conjugate to ${\bf B}_{\mathrm{tot}}$.

\subsection{Notation and setup}
\label{sec:multidim_setup}
For a multiindex $\mi{a}=( a_{1},\dots, a_{p})\in\mathbb{N}_{0}^{p}$
we use multi-index notation (see e.g.~\cite{constantine1996multivariate})
\begin{equation}
\label{eq:multiindex_conv}
\begin{aligned}
&|\mi{a}|\equiv\sum_{i=1}^{p} a_{i},\quad
\mi{a}!\equiv\prod_{i=1}^{p} a_{i}!,\quad
{\bf u}^{\mi{a}}\equiv\prod_{i=1}^{p}u_{i}^{ a_{i}},\quad
\partial_{{\bf u}}^{\mi{a}}\equiv\prod_{i=1}^{p}\frac{\partial^{ a_{i}}}{\partial u_{i}^{ a_{i}}}, \quad
\binom{\mi{a}}{\mi{b}}\equiv\frac{\mi{a}!}{\mi{b}!\,(\mi{a}-\mi{b})!},\\
&\mi{b}\le\mi{a}\iff b_{i}\le  a_{i}\ \forall i, \qquad
\mi{e}_j \equiv (0,\ldots,\, \underbrace{1}_{\text{index}~j} \,,\ldots,0).
\qquad
\mi{0}\equiv(0,\dots,0).
\end{aligned}
\end{equation}

The unconstrained (reference) ensemble is specified by the joint CGF
\begin{equation}
\label{eq:CGF_multidim}
G^{\mathrm{gce}}({\bf t},\boldsymbol{\lambda})
\equiv
\ln\Big\langle e^{{\bf t}\cdot{\bf X}+\boldsymbol{\lambda}\cdot{\bf B}_{\mathrm{tot}}}\Big\rangle_{\mathrm{gce}}
=
\sum_{\mi{n}\ge\mi{0}}\ \sum_{\mi{m}\ge\mi{0}}
\frac{\kappa^{\bf XB}_{\mi{n}\,\mi{m}}}{\mi{n}!\,\mi{m}!}\,
{\bf t}^{\mi{n}}\boldsymbol{\lambda}^{\mi{m}}, \qquad \mi{n} \in \mathbb{N}_0^d, \qquad \mi{m} \in \mathbb{N}_0^s,
\end{equation}
which defines the mixed grand-canonical cumulants $\kappa^{\bf XB}_{\mi{n}\,\mi{m}}$.
In the one-dimensional case of Sec.~\ref{sec:formalism} ($d=s=1$), the multi-indices reduce to scalars and one recovers $\kappa^{\bf XB}_{\mi{n}\,\mi{m}}=\kappa^{XB}_{n\,m}$. For multiple named observables and conserved charges (e.g.\ two observables $X,Y$ and two charges $B,Q$) the explicit charge-labeled form $\kappa^{XYBQ}_{a_1a_2b_1b_2}$ is equivalent to $\kappa^{\bf XB}_{\mi{a}\,\mi{b}}$ with $\mi{a}=(a_1,a_2)$ and $\mi{b}=(b_1,b_2)$.
The explicit charge-labeled formulas for the QCD conserved charges $B$, $Q$, $S$ are given in Sec.~\ref{sec:qcd_charges}.

Exact global conservation is enforced by a product of Kronecker constraints for each charge,
$\delta_{{\bf B}_{\rm tot},{\bf B}_0}\equiv\prod_{b=1}^{s}\delta_{B^{b}_{\rm tot},B^{b}_0}$,
and the canonical CGF $G^{\rm ce}({\bf t})\equiv\ln\langle e^{{\bf t}\cdot{\bf X}}\rangle_{\rm ce}$
has the same conditional-probability structure as Eq.~\eqref{eq:G_ce} with $X\!\to\!{\bf X}$ and $B_{\rm tot}\!\to\!{\bf B}_{\rm tot}$.
Compared to Sec.~\ref{sec:formalism}, the only structural change is that the Fourier representation of
$\delta_{{\bf B}_{\rm tot},{\bf B}_0}$ introduces an $s$-dimensional integral and an $s$-component saddle variable.
The contour-deformation and saddle-uniqueness arguments of Sec.~\ref{sec:saddlepoint} can be repeated in the multidimensional case. For real $\mathbf t$, strict convexity in real $\boldsymbol\lambda$ is now guaranteed by the Hessian $\nabla^2_{\boldsymbol{\lambda}}G^{\rm gce}$, which is the covariance matrix of ${\bf B}_{\rm tot}$ in the reweighted ensemble and is positive definite for a non-degenerate charge basis.

In the leading-order saddle-point approximation (thermodynamic limit), the multidimensional analogue of Eq.~(\ref{eq:Gce_final}) reads
\begin{equation}
\label{eq:CGF_ce_multidim_saddle}
G^{\mathrm{ce}}({\bf t})
\simeq
G^{\mathrm{gce}}\!\big[{\bf t},\boldsymbol{\lambda}({\bf t})\big]
-\boldsymbol{\lambda}({\bf t})\!\cdot\!{\bf B}_{0},
\qquad
\nabla_{\boldsymbol{\lambda}}\,G^{\mathrm{gce}}\!\big[{\bf t},\boldsymbol{\lambda}({\bf t})\big]={\bf B}_{0},
\qquad
\boldsymbol{\lambda}({\bf 0})={\bf 0}.
\end{equation}
Here we have neglected the subleading Gaussian prefactor corrections
to the canonical CGF (see Appendix~\ref{sec:finite_size}).
Following Sec.~\ref{sec:formalism}, the bare symbol $\boldsymbol{\lambda}$ denotes the free source variable of $G^{\rm gce}(\mathbf{t},\boldsymbol{\lambda})$ in Eq.~\eqref{eq:CGF_multidim}, while $\boldsymbol{\lambda}(\mathbf{t})$ denotes the saddle-point solution of $\nabla_{\boldsymbol{\lambda}}G^{\rm gce}=\mathbf{B}_0$, the multidimensional analogue of $\lambda_\star(t)$.
Unlike the one-dimensional case, the saddle condition is a system of $s$ coupled equations.
We introduce the $s\times s$ GCE charge covariance matrix
\begin{align}
\label{eq:Kab}
K_{ab}\equiv
\partial_{\lambda_a}\partial_{\lambda_b}G^{\rm gce}({\bf 0},{\bf 0})
=\kappa^{\bf B}_{\mi{e}_a+\mi{e}_b}
\equiv {\rm cov}_{\rm gce}(B_a,B_b),
\qquad a,b=1,\dots,s,
\end{align}
and assume non-degeneracy, ${\rm det}\,\bf K\neq 0$, so that the implicit system for $\boldsymbol{\lambda}({\bf t})$ can be solved locally.\footnote{If ${\bf K}$ is singular,
for example when the chosen charge basis is linearly dependent (such as $(B,Q,B+Q)$) or includes a charge that does not fluctuate in the reference ensemble,
one must reduce to an independent fluctuating charge basis before inversion.
}

Setting $B_0^a = \kappa^{\bf B}_{\mi{e}_a}$ (the GCE mean, as in Sec.~\ref{sec:formalism}) ensures $\boldsymbol{\lambda}({\bf 0})={\bf 0}$, so each $\lambda_a({\bf t})$ admits a formal power-series expansion in ${\bf t}$:
\begin{equation}
  \label{eq:lambda_series_multi}
  \lambda_a({\bf t})
   = \sum_{\mi{n} \geq \mi{0}}
  \dfrac{\lambda_{a;\mi{n}}}{\mi{n}!}\,{\bf t}^\mi{n},
  \qquad
  \lambda_{a;\mi{n}} \equiv \partial_{\bf t}^{\mi{n}}\lambda_a|_{{\bf t}=\mi{0}}, \qquad a=1,\dots,s, \qquad \mi{n} \in \mathbb{N}_0^d.
\end{equation}
The Taylor coefficients are collected into the vector $\boldsymbol{\lambda}_{\mi{n}}\equiv(\lambda_{1;\mi{n}},\dots,\lambda_{s;\mi{n}})\in\mathbb{R}^s$, used below as the argument of the multivariate Bell polynomials [Eq.~\eqref{eq:Bell_multi_explicit}].
The key identity
\begin{equation}
\partial_{{\bf t}}^{\mi{n}}\partial_{\boldsymbol{\lambda}}^{\mi{m}+\mi{e}_a}
G^{\rm gce}({\bf t},\boldsymbol{\lambda})\Big|_{({\bf 0},{\bf 0})}
=\kappa^{\bf XB}_{\mi{n},\mi{m}+\mi{e}_a},
\label{eq:kappa_as_derivs}
\end{equation}
allows the saddle condition $\nabla_{\boldsymbol{\lambda}}G^{\rm gce}({\bf t},\boldsymbol{\lambda}({\bf t}))={\bf B}_0$ to be expanded order by order in ${\bf t}$ (cf.~Sec.~\ref{sec:formalism}), yielding via the multivariate Fa\`a di Bruno formula the linear system for $\boldsymbol{\lambda}_\mi{n}$ derived below.

\subsection{Multivariate Fa\`a di Bruno formula}
\label{sec:multidim_fdb}

We define the multidimensional saddle-point condition function
\eq{
S_a({\bf t}) \equiv \partial_{\lambda_a} G^{\rm gce}({\bf t},\boldsymbol{\lambda}({\bf t})) - B_0^a,
}
which is an analogue of Eq.~\eqref{eq:Sonedim} in Sec.~\ref{sec:formalism} for the one-dimensional case.
Applying the multivariate Fa\`a di Bruno formula~\cite{constantine1996multivariate,leipnik2007multivariate} and using Eq.~\eqref{eq:kappa_as_derivs} one can write the derivative $\partial_{{\bf t}}^{\mi{n}}\equiv\partial_{t_1}^{n_1}\cdots\partial_{t_d}^{n_d}$, with $\mi{n}=(n_1,\dots,n_d)\in\mathbb{N}_0^d$, of $S$ for any multiindex $\mi{n}$ as follows:
\begin{equation}
\partial_{{\bf t}}^{\mi{n}} S_a({\bf t})\Big|_{{\bf 0}}
=
\kappa^{\bf XB}_{\mi{n}\,\mi{e}_a}
+\sum_{{\bf 0}\leq\mi{p}\leq \mi{n}}\binom{\mi{n}}{\mi{p}}
\sum_{1\leq|\mi{q}|\leq|\mi{n}-\mi{p}|}
\kappa^{\bf XB}_{\mi{p},\mi{q}+\mi{e}_a}\;
\mathcal{B}_{\mi{n}-\mi{p},\,\mi{q}}
\!\left(\{\boldsymbol{\lambda}_{\mi{k}}\}\right),
\label{eq:Sder_multi}
\end{equation}
where $\boldsymbol{\lambda}_{\mi{k}}\equiv\partial_{{\bf t}}^{\mi{k}}\boldsymbol{\lambda}({\bf t})|_{{\bf 0}}$
and $\mathcal{B}_{\mi{r},\mi{q}}$ denote the multivariate analogues of the partial exponential Bell polynomials~\cite{Comtet:1974,Withers:2008bell,Schumann:2019bell}.

We define $\mathcal{B}_{\mi{r},\mi{q}}$ as a sum over colored set partitions (see e.g.\ Ref.~\cite{Vovchenko:2021yen}):
\begin{equation}
\mathcal{B}_{\mi{r},\mi{q}}
\!\left(\{\boldsymbol{\lambda}_{\mi{k}}\}\right)
 =
\sum_{\substack{\pi\in\Pi_{|\mi{r}|}^{(s)}\\ \mi{q}(\pi)=\mi{q}}}
\prod_{j=1}^s\ \prod_{\blk\in\pi_j}\ \lambda_{j;\mi{k}_\blk}.
\label{eq:Bell_multi_explicit}
\end{equation}
where $\mi{r}\in\mathbb{N}_0^d$ encodes the derivative directions of $\partial_{\bf t}^\mi{r}$ being partitioned, while $\mi{q}\in\mathbb{N}_0^s$ counts the blocks of each color.
Here $\Pi_{N}^{(s)}$ ($N\equiv|\mi{r}|$) denotes the set of all possible partitions~\cite{haiman1989incidence,constantine1996multivariate} of a set of $N$ elements into nonempty colored blocks, where each block may carry a color $j\in\{1,\dots,s\}$.
For $\pi\in\Pi_N^{(s)}$ we denote by $\pi_j$ the collection (set) of blocks of color $j$; the multiindex $\mi{q}(\pi)=(|\pi_1|,\dots,|\pi_s|)\in\mathbb{N}_0^s$ then counts the number of blocks of each color, with $|\pi|=\sum_{j=1}^s|\pi_j|=|\mi{q}(\pi)|$ being the total number of blocks.
The constraint $\mi{q}(\pi)=\mi{q}$ fixes the number of blocks of each color, i.e. only this subset of all partitions is included in the sum.
The multiindex $\mi{k}_{\blk}\in\mathbb{N}_0^d$ records which derivative directions are grouped in block~$\blk$.
Concretely, the operator $\partial_{\bf t}^{\mi{r}}$ comprises $|\mi{r}|$ first-order derivatives $\partial_{t_i}$
(with $r_i$ copies of $\partial_{t_i}$ for each $i=1,\dots,d$);
a partition $\pi$ distributes these $|\mi{r}|$ derivatives among its blocks, and the component $(\mi{k}_{\blk})_i$ counts how many $\partial_{t_i}$'s are assigned to block~$\blk$.
One therefore has $|\mi{k}_{\blk}|=|\blk|$ and $\sum_{\blk\in\pi}\mi{k}_{\blk}=\mi{r}$.

As a concrete illustration, take $d=1$ (so $\mi{k}_\blk=|\blk|$) and $s=2$, with $\mi{r}=3$ and $\mi{q}=(1,1)$ (one block per color). The set partitions of $\{1,2,3\}$ into two non-empty blocks are $\{\{1\},\{2,3\}\}$, $\{\{2\},\{1,3\}\}$, $\{\{3\},\{1,2\}\}$; each admits two color assignments (the size-1 block in color~$1$ or in color~$2$). Summing the six colored contributions,
\begin{equation}
\mathcal{B}_{3,(1,1)}\!\left(\{\boldsymbol{\lambda}_{\mi{k}}\}\right) = 3\,\lambda_{1;1}\lambda_{2;2}+3\,\lambda_{1;2}\lambda_{2;1}.
\end{equation}

From Eq.~\eqref{eq:Bell_multi_explicit} one sees that, for $|\mi{r}|\geq 1$, $\mathcal{B}_{\mi{r},\mi{q}}=0$ unless $1\leq |\mi{q}|\leq |\mi{r}|$.

A key property of the multivariate Bell polynomials is that for $\mi{q}=\mi{e}_j$ and $\mi{n}\neq \mi{0}$ one has 
\begin{equation}
\mathcal{B}_{\mi{n},\,{\bf e}_j}\!\left(\{\boldsymbol{\lambda}_{\mi{k}}\}\right)
=\lambda_{j;\mi{n}},
\qquad j=1,\dots,s,
\label{eq:bell_key_property}
\end{equation}
where $\lambda_{j;\mi{n}}$ denotes the $j$-th component of the vector coefficient
$\boldsymbol{\lambda}_{\mi{n}}=\partial_{{\bf t}}^{\mi{n}}\boldsymbol{\lambda}({\bf t})|_{{\bf 0}}$.
Moreover, $\lambda_{j;\mi{n}}$ can arise in \eqref{eq:Sder_multi} only from the unique choice
$\mi{p}=\mi{0}$ and $\mi{q}=\mi{e}_j$.

Separating these contributions via Eq.~\eqref{eq:bell_key_property}
and imposing the constraint $S_a({\bf t})\equiv 0$ (hence $\partial_{{\bf t}}^{\mi{n}}S_a|_{{\bf 0}}=0$ for $\mi{n}\neq\mi{0}$) yields a linear system
\begin{equation}
0=
\kappa^{\bf XB}_{\mi{n}\,\mi{e}_a}
+\sum_{b=1}^s K_{ab}\,\lambda_{b;\mi{n}}
+R_{a,\mi{n}},
\qquad a=1,\dots,s, \quad \mi{n}\neq \mi{0}.
\label{eq:lambda_linear_system}
\end{equation}
which is used to determine $\boldsymbol{\lambda}_{\mi{n}}=(\lambda_{1;\mi{n}},\dots,\lambda_{s;\mi{n}})$ iteratively in increasing total degree $|\mi{n}|$. Here $K_{ab}$ is the charge cumulant matrix \eqref{eq:Kab}.

The remainder $R_{a,\mi{n}}$ contains only coefficients $\boldsymbol{\lambda}_{\mi{k}}$ of strictly lower total degree,
$0<|\mi{k}|<|\mi{n}|$, and is given explicitly by 
\begin{equation}
R_{a,\mi{n}}=
\sum_{{\mi 0}\leq\mi{p}\leq \mi{n}}\binom{\mi{n}}{\mi{p}}
\sum_{\mi{q}\in \mathbb{N}_0^s}^{\prime}\
\kappa^{\bf XB}_{\mi{p},\mi{q}+\mi{e}_a}\;
\mathcal{B}_{\mi{n}-\mi{p},\,\mi{q}}\!\left(\{\boldsymbol{\lambda}_{\mi{k}}\}\right).
\label{eq:R_def}
\end{equation}

Here $\displaystyle \sum_{\mi{q}\in \mathbb{N}_0^s}^{\prime}$ denotes sum over $\mi{q}$ with  $1\leq|\mi{q}|\leq|\mi{n}-\mi{p}|$ excluding $\mi{q}=\mi{e}_j$ when $\mi{p}={\bf 0}$.

Assuming matrix ${\bf K}$ is invertible
we obtain the explicit recursion
\begin{equation}
\lambda_{j;\mi{n}}
=
-\sum_{a=1}^s (K^{-1})_{ja}\,\Bigl(\kappa^{\bf XB}_{\mi{n}\,\mi{e}_a}+R_{a,\mi{n}}\Bigr),
\qquad
\mi{n} \neq \mi{0},\quad {\rm det}\, {\bf K} \neq 0
\label{eq:lambda_recursion_explicit}
\end{equation}
which is the multi-dimensional analogue of the one-dimensional formula \eqref{eq:lambda_n}. 

Using Eq.~\eqref{eq:Bell_multi_explicit} the remainder can be written as
\begin{align}
R_{a,\mi{n}}
=
\sum_{{\bf 0}\leq\mi{p}\leq \mi{n}}
\binom{\mi{n}}{\mi{p}}
\sum_{
\pi\in\Pi_{|\mi{n}-\mi{p}|}^{(s)}
}^{\prime}\
\kappa^{\bf XB}_{\mi{p},\mi{q}(\pi)+\mi{e}_a}\;
\prod_{j=1}^s\ \prod_{\blk\in\pi_j}\ \lambda_{j;\mi{k}_{\blk}} .
\label{eq:R_colored_def}
\end{align}
Here prime over the sum means $1\leq|\pi|\leq|\mi{n}-\mi{p}|$ and 
$(\mi{p}=\mi{0})\Rightarrow(|\pi|\geq 2)$. Therefore in the sum we omit the contributions with $\mi{p}=\mi{0}$ and $|\pi|=1$
(equivalently, $\mi{q}(\pi)=\mi{e}_j$ for some $j=1,\dots,s$), which are precisely the terms
$\mathcal{B}_{\mi{n},\,{\bf e}_j}=\lambda_{j;\mi{n}}$ isolated on the l.h.s. of
Eq.~\eqref{eq:lambda_recursion_explicit}.

\subsection{Iterative solution for canonical cumulants}
\label{sec:multidim_kappa_ce}

Differentiating the saddle-point expression~\eqref{eq:CGF_ce_multidim_saddle} for cumulant-generating function $G^{\rm ce}(\bf t)$ with respect to ${\bf t}$ yields the canonical cumulants,
\eq{
\kappa^{{\bf X}|{\bf B}}_{\mi{n}}=\partial_{{\bf t}}^{\mi{n}}G^{\rm gce}({\bf t},\boldsymbol{\lambda}({\bf t}))|_{{\bf 0}}-{\bf B}_0\cdot\boldsymbol{\lambda}_{\mi{n}},
}
where $\kappa^{{\bf X}|{\bf B}}_{\mi{n}}$ is the joint cumulant involving $n_i$-th order of the $i$-th observable $X_i$.
The first term is the $\mi{n}$-th derivative of the composite function $G^{\rm gce}$ evaluated along the saddle-point path $\boldsymbol{\lambda}({\bf t})$, and the second term arises from the $-\boldsymbol{\lambda}({\bf t})\cdot{\bf B}_0$ piece.
Applying the multivariate Fa\`a di Bruno formula to the first term,
for any multiindex $\mi{n}\in\mathbb{N}_0^d$ one finds
\begin{equation}
\label{eq:ggce-fdB-multidim}
\partial_{{\bf t}}^{\mi{n}}
G^{\rm gce}({\bf t},\boldsymbol{\lambda}({\bf t}))\Big|_{{\bf t}={\bf 0}}
=
\sum_{\mi{0}\le\mi{p}\le \mi{n}}
\binom{\mi{n}}{\mi{p}}
\sum_{\substack{\mi{q}\in\mathbb{N}_0^s\\ 0\le |\mi{q}|\le |\mi{n}-\mi{p}|}}
\kappa^{\bf XB}_{\mi{p}\,\mi{q}}\;
\mathcal{B}_{\mi{n}-\mi{p},\,\mi{q}}\!\left(\{\boldsymbol{\lambda}_{\mi{k}}\}\right).
\end{equation}
Here $\mi{p}\in\mathbb{N}_0^d$ labels the direct ${\bf t}$-derivatives and $\mi{q}\in\mathbb{N}_0^s$ labels the $\boldsymbol{\lambda}$-derivatives,
$\kappa^{\bf XB}_{\mi{p}\,\mi{q}}$ are the mixed GCE cumulants defined via Eq.~\eqref{eq:CGF_multidim},
and $\mathcal{B}_{\mi{n}-\mi{p},\,\mi{q}}$ are the multivariate Bell polynomials of Eq.~\eqref{eq:Bell_multi_explicit} evaluated at the saddle-point coefficients $\{\boldsymbol{\lambda}_{\mi{k}}\}$.

By combining Eqs.~(\ref{eq:CGF_ce_multidim_saddle}) and (\ref{eq:ggce-fdB-multidim}) we obtain, for any multiindex $\mi{n}$ with $|\mi{n}|\ge 1$,
\begin{align}
\label{eq:kce-multidim-master}
\kappa^{{\bf X}|{\bf B}}_{\mi{n}}
& =
\kappa^{\bf X}_{\mi{n}}
+
\sum_{\mi{0}\le\mi{p}\le \mi{n}}
\binom{\mi{n}}{\mi{p}}
\sum_{\substack{\mi{q}\in\mathbb{N}_0^s\\ 1\le |\mi{q}|\le |\mi{n}-\mi{p}|}}
\kappa^{\bf XB}_{\mi{p}\,\mi{q}}\;
\mathcal{B}_{\mi{n}-\mi{p},\,\mi{q}}\!\left(\{\boldsymbol{\lambda}_{\mi{k}}\}\right)
\;-\;
{\bf B}_0\cdot \boldsymbol{\lambda}_{\mi{n}}\nonumber\\
& = 
\kappa^{\bf X}_{\mi{n}}
+
\sum_{\mi{0}\le\mi{p}\le \mi{n}}
\binom{\mi{n}}{\mi{p}}
\sum_{\substack{\pi\in \Pi^{(s)}_{|\mi{n}-\mi{p}|}\\ 1\le |\pi|\le |\mi{n}-\mi{p}|}}
\kappa^{\bf XB}_{\mi{p}\,\mi{q}(\pi)}\;
\prod_{j=1}^{s}\;\prod_{\blk\in\pi_j}\lambda_{j;\mi{k}_{\blk}}
\;-\;
{\bf B}_0\cdot \boldsymbol{\lambda}_{\mi{n}}
\nonumber\\
& =
\kappa^{\bf X}_{\mi{n}}
+
\sum_{\mi{0}\le\mi{p}\le \mi{n}}
\binom{\mi{n}}{\mi{p}}
\sum_{\substack{\pi\in \Pi^{(s)}_{|\mi{n}-\mi{p}|}\\ 1\le |\pi|\le |\mi{n}-\mi{p}|}}^{\prime\prime}\
\kappa^{\bf XB}_{\mi{p}\,\mi{q}(\pi)}\;
\prod_{j=1}^{s}\;\prod_{\blk\in\pi_j}\lambda_{j;\mi{k}_{\blk}}\,,
\end{align}
where $\sum_\pi^{\prime\prime}$ additionally excludes terms with $\mi{p}=\mi{0}$ and $|\pi|=1$, i.e.\ $\mi{q}(\pi)=\mi{e}_j$ for some $j$; summed over $j$, these yield $\sum_j B_0^j\lambda_{j;\mi{n}}={\bf B}_0\cdot\boldsymbol{\lambda}_\mi{n}$, cancelling the last term (cf.\ the analogous cancellation in Eq.~\eqref{eq:kappa-ce-general}).
Together with the recursion for $\boldsymbol{\lambda}_{\mi{n}}$ given by Eqs.~(\ref{eq:lambda_linear_system})--(\ref{eq:lambda_recursion_explicit}),
Eq.~(\ref{eq:kce-multidim-master}) provides an iterative procedure to compute $\kappa^{{\bf X}|{\bf B}}_{\mi{n}}$ order-by-order
in $|\mi{n}|$.

We list explicit expressions up to third order (with $i,j,k\in\{1,\dots,d\}$ labeling the in-acceptance observables and $a,b,c\in\{1,\dots,s\}$ the components of $\mathbf{B}_{\rm tot}$; we write $X_i$ and $B_a$ as shorthands for $X^i_{\rm in}$ and $B^a_{\rm tot}$ to keep the formulas compact, and cumulants are symmetric under permutations of indices of the same kind):
\begin{align}
\label{eq:kce1-general}
\kappa(X_i | {\bf B})
&=
\kappa(X_i),
\\[2mm]\label{eq:kce2-general}
\kappa(X_i, X_j | {\bf B})
&=
\kappa(X_i, X_j)
-\kappa(X_i, B_a)\,(K^{-1})^{ab}\,\kappa(X_j, B_b)
\nonumber\\&
=
\kappa(X_i, X_j)
-\kappa(X_i, B_a)\,A_j^{a},
\\[2mm]\label{eq:kce3-general}
\kappa(X_i, X_j, X_k | {\bf B})
&=
\kappa(X_i, X_j, X_k)
-\kappa(X_i, X_j, B_a)\,A_k^{a}
-\kappa(X_i, X_k, B_a)\,A_j^{a}
\nonumber\\&
-\kappa(X_j, X_k, B_a)\,A_i^{a}
+\kappa(X_i, B_a, B_b)\,A_j^{a}A_k^{b}
\nonumber\\&
+\kappa(X_j, B_a, B_b)\,A_i^{a}A_k^{b}
+\kappa(X_k, B_a, B_b)\,A_i^{a}A_j^{b}
\nonumber\\&
-\kappa(B_a, B_b, B_c)\,A_i^{a}A_j^{b}A_k^{c},
\end{align}
with implicit summation over repeated charge indices $a,b,c=1,\dots,s$.
Here $K_{ab}$ is the charge cumulant matrix of Eq.~\eqref{eq:Kab},
and we introduce the linear-response coefficient
\eq{
\label{eq:A-def}
A_i^{a}\equiv (K^{-1})^{ab}\,\kappa(X_i, B_b),
}
which is the covariance of $X_i$ with $B_a$ normalized by the charge covariance matrix.\footnote{$A_i^{a}$ is the linear regression coefficient of $X_i$ on $B_a$ in the unconstrained reference ensemble, quantifying how strongly the observable $X_i$ couples to the conserved charge $B_a$.} In the binomial-acceptance and SAM-1.0 limits it reduces to the scalar acceptance fraction, $A_i^{a}\to\alpha\,\delta_i^{a}$ (cf.\ Secs.~\ref{sec:sam-1},~\ref{sec:binomial_limit}); more generally it generalizes $\alpha$ to arbitrary observables (including non-conserved quantities such as net protons), arbitrary joint distributions $P_{\rm gce}({\bf X},{\bf B}_{\rm tot})$ (including acceptances with non-trivial in/out correlations), and multiple simultaneously conserved charges.
Here $(K^{-1})^{ab}$ denotes the $(a,b)$ entry of the inverse of the matrix~$\mathbf{K}$ and can be expressed via Cramer's rule as $(K^{-1})^{ab} = (-1)^{a+b}\det\mathbf{K}^{(b,a)}/\det\mathbf{K}$, where $\mathbf{K}^{(b,a)}$ is the submatrix obtained by deleting row $b$ and column $a$.

We also list the fourth-order cumulant, 
\begin{align}
&\kappa(X_i, X_j, X_k, X_l | {\bf B})
=
\kappa(X_i, X_j, X_k, X_l)
-\kappa(X_i, X_j, X_k, B_a)\,A_l^{a}
\nonumber\\&\quad
-\kappa(X_i, X_j, X_l, B_a)\,A_k^{a}
-\kappa(X_i, X_k, X_l, B_a)\,A_j^{a}
-\kappa(X_j, X_k, X_l, B_a)\,A_i^{a}
\nonumber\\&\quad
+\kappa(X_i, X_j, B_a, B_b)\,A_k^{a}A_l^{b}
+\kappa(X_i, X_k, B_a, B_b)\,A_j^{a}A_l^{b}
+\kappa(X_i, X_l, B_a, B_b)\,A_j^{a}A_k^{b}
\nonumber\\&\quad
+\kappa(X_j, X_k, B_a, B_b)\,A_i^{a}A_l^{b}
+\kappa(X_j, X_l, B_a, B_b)\,A_i^{a}A_k^{b}
+\kappa(X_k, X_l, B_a, B_b)\,A_i^{a}A_j^{b}
\nonumber\\&\quad
-\kappa(X_i, B_a, B_b, B_c)\,A_j^{a}A_k^{b}A_l^{c}
-\kappa(X_j, B_a, B_b, B_c)\,A_i^{a}A_k^{b}A_l^{c}
\nonumber\\&\quad
-\kappa(X_k, B_a, B_b, B_c)\,A_i^{a}A_j^{b}A_l^{c}
-\kappa(X_l, B_a, B_b, B_c)\,A_i^{a}A_j^{b}A_k^{c}
\nonumber\\&\quad
+\kappa(B_a, B_b, B_c, B_d)\,A_i^{a}A_j^{b}A_k^{c}A_l^{d}
\nonumber\\&\quad
-U_a^{ij}\,(K^{-1})^{ab}\,U_b^{kl}
-U_a^{ik}\,(K^{-1})^{ab}\,U_b^{jl}
-U_a^{il}\,(K^{-1})^{ab}\,U_b^{jk},
\label{eq:kce4_general}
\end{align}
where the last line contains the connected-correction terms with
\eq{
U_a^{ij} \equiv \kappa(X_i, X_j, B_a)
-\kappa(X_i, B_a, B_b)\,A_j^{b}
-\kappa(X_j, B_a, B_b)\,A_i^{b}
+\kappa(B_a, B_b, B_c)\,A_i^{b}A_j^{c}.
\label{eq:Uaij_def}
}

In the diagonal case of a single observable ($d=1$), all indices $i,j,k,\dots$ coincide.
Correspondingly, $A_i^{a}\ \to\ A^{a}\equiv(K^{-1})^{ab}\,\kappa(X, B_b),$
so that the canonical cumulants for an arbitrary number of conserved charges $s$ read
\begin{align}
\kappa^{X|{\bf B}}_1&=\kappa^{X}_1,\label{eq:k1-diag}
\\[2mm]
\kappa^{X|{\bf B}}_2&=\kappa^{X}_2-\kappa(X, B_a)A^{a},\label{eq:k2-diag}
\\[2mm]
\kappa^{X|{\bf B}}_3&=\kappa^{X}_3
-3\,\kappa(X, X, B_a)A^{a}
+3\,\kappa(X, B_a, B_b)A^{a}A^{b}
-\kappa(B_a, B_b, B_c)A^{a}A^{b}A^{c},\label{eq:k3-diag}
\\[2mm]
\kappa^{X|{\bf B}}_4&=\ \kappa^{X}_4
-4\,\kappa(X, X, X, B_a)\,A^{a}
+6\,\kappa(X, X, B_a, B_b)\,A^{a}A^{b}
\nonumber\\&
-4\,\kappa(X, B_a, B_b, B_c)\,A^{a}A^{b}A^{c}
+\kappa(B_a, B_b, B_c, B_d)\,A^{a}A^{b}A^{c}A^{d}
\nonumber\\&
-3\,U_{a}\,(K^{-1})^{ab}\,U_{b}\,,\label{eq:k4-diag}
\end{align}
with
\begin{equation}
U_{a}\equiv \kappa(X, X, B_a)-2\,\kappa(X, B_a, B_b)\,A^{b}
+\kappa(B_a, B_b, B_c)\,A^{b}A^{c}.
\label{eq:Ua_def}
\end{equation}

\subsection{QCD conserved charges}
\label{sec:qcd_charges}

We now consider the specific case of QCD conserved charges: baryon number $B$, electric charge $Q$, and strangeness $S$.
In all cases, the mean at leading order is unaffected by conservation, $\kappa^{X|{\bf B}}_1 = \kappa^{X}_{1}$.

In the case of a single conserved charge, i.e., $s=1$ (${\bf B}=B$), the diagonal cumulants reduce to Eqs.~(\ref{eq:k1ce})--(\ref{eq:k4ce}).

For mixed cumulants involving two observables $X$ and $Y$, the results up to third order read
\begin{align}
\kappa^{XY|B}_{1\,1}
&=\kappa^{XY}_{1\,1}-\frac{\kappa^{XB}_{1\,1}\,\kappa^{YB}_{1\,1}}{\kappa^{B}_{2}} \,, \label{eq:k11SAMB}
\\[2mm]
\kappa^{XY|B}_{2\,1}
&=\kappa^{XY}_{2\,1}
-\frac{2\,\kappa^{XB}_{1\,1}\,\kappa^{XYB}_{1\,1\,1}+\kappa^{XB}_{2\,1}\,\kappa^{YB}_{1\,1}}{\kappa^{B}_{2}}
\nonumber\\&\quad
+\frac{2\,\kappa^{XB}_{1\,1}\,\kappa^{XB}_{1\,2}\,\kappa^{YB}_{1\,1}+\left(\kappa^{XB}_{1\,1}\right)^{2}\kappa^{YB}_{1\,2}}{\left(\kappa^{B}_{2}\right)^{2}}
-\frac{\kappa^{B}_{3}\,\left(\kappa^{XB}_{1\,1}\right)^{2}\kappa^{YB}_{1\,1}}{\left(\kappa^{B}_{2}\right)^{3}} \,.
\end{align}

In the case of two conserved charges ($s=2$), e.g., ${\bf B}=\{B,Q\}$, up to second order one obtains
\begin{align}
\label{eq:k2s2-explicit}
\kappa^{X|BQ}_{2} &=
\kappa^{X}_{2}
-\frac{
\kappa^{Q}_{2}\left(\kappa^{XB}_{1\,1}\right)^{2}
-2\,\kappa^{BQ}_{1\,1}\,\kappa^{XB}_{1\,1}\,\kappa^{XQ}_{1\,1}
+\kappa^{B}_{2}\left(\kappa^{XQ}_{1\,1}\right)^{2}
}{
{\rm det}\,{\bf K}}\,,\\[2mm]
\kappa^{XY|BQ}_{1\,1}&=
\kappa^{XY}_{1\,1}
-\frac{
\kappa^{Q}_{2}\,\kappa^{XB}_{1\,1}\,\kappa^{YB}_{1\,1}
+\kappa^{B}_{2}\,\kappa^{XQ}_{1\,1}\,\kappa^{YQ}_{1\,1}
-\kappa^{BQ}_{1\,1}
\Big(\kappa^{XB}_{1\,1}\,\kappa^{YQ}_{1\,1}
+\kappa^{XQ}_{1\,1}\,\kappa^{YB}_{1\,1}\Big)
}{
{\rm det}\,{\bf K}
}\,,
\end{align}
where ${\rm det}\,{\bf K}=\kappa^{B}_{2}\,\kappa^{Q}_{2}-\left(\kappa^{BQ}_{1\,1}\right)^{2}$.

For three conserved charges ($s=3$), e.g., ${\bf B}=\{B,Q,S\}$, the determinant of the $3\times 3$ charge cumulant matrix~\eqref{eq:Kab} reads
\begin{align}
\label{eq:detK3}
{\rm det}\,{\bf K} &=
\kappa^{B}_{2}\,\kappa^{Q}_{2}\,\kappa^{S}_{2}
+2\,\kappa^{BQ}_{1\,1}\,\kappa^{BS}_{1\,1}\,\kappa^{QS}_{1\,1}
-\left(\kappa^{BQ}_{1\,1}\right)^{2}\kappa^{S}_{2}
-\left(\kappa^{QS}_{1\,1}\right)^{2}\kappa^{B}_{2}
-\left(\kappa^{BS}_{1\,1}\right)^{2}\kappa^{Q}_{2}\,,
\end{align}
and the second-order results are
\begin{align}
\label{eq:k2s3-explicit}
\kappa^{X|BQS}_{2} &=\kappa^{X}_{2}
-\frac{1}{{\rm det}\,{\bf K}}\biggl\{
\left(\kappa^{XB}_{1\,1}\right)^{2}
\Big(\kappa^{Q}_{2}\,\kappa^{S}_{2}-\left(\kappa^{QS}_{1\,1}\right)^{2}\Big)
+\left(\kappa^{XQ}_{1\,1}\right)^{2}
\Big(\kappa^{B}_{2}\,\kappa^{S}_{2}-\left(\kappa^{BS}_{1\,1}\right)^{2}\Big)
\nonumber\\[1mm]&\qquad
+\left(\kappa^{XS}_{1\,1}\right)^{2}
\Big(\kappa^{B}_{2}\,\kappa^{Q}_{2}-\left(\kappa^{BQ}_{1\,1}\right)^{2}\Big)
+2\,\kappa^{XB}_{1\,1}\,\kappa^{XQ}_{1\,1}
\Big(\kappa^{BS}_{1\,1}\,\kappa^{QS}_{1\,1}-\kappa^{BQ}_{1\,1}\,\kappa^{S}_{2}\Big)
\nonumber\\[1mm]&\qquad
+2\,\kappa^{XB}_{1\,1}\,\kappa^{XS}_{1\,1}
\Big(\kappa^{BQ}_{1\,1}\,\kappa^{QS}_{1\,1}-\kappa^{Q}_{2}\,\kappa^{BS}_{1\,1}\Big)
+2\,\kappa^{XQ}_{1\,1}\,\kappa^{XS}_{1\,1}
\Big(\kappa^{BQ}_{1\,1}\,\kappa^{BS}_{1\,1}-\kappa^{B}_{2}\,\kappa^{QS}_{1\,1}\Big)
\biggr\}\,,
\\[3mm]
\label{eq:k11s3-explicit}
\kappa^{XY|BQS}_{1\,1}&=
\kappa^{XY}_{1\,1}
-\frac{1}{{\rm det}\,{\bf K}}\biggl\{
\kappa^{XB}_{1\,1}\,\kappa^{YB}_{1\,1}
\Big(\kappa^{Q}_{2}\,\kappa^{S}_{2}-\left(\kappa^{QS}_{1\,1}\right)^{2}\Big)
+\kappa^{XQ}_{1\,1}\,\kappa^{YQ}_{1\,1}
\Big(\kappa^{B}_{2}\,\kappa^{S}_{2}-\left(\kappa^{BS}_{1\,1}\right)^{2}\Big)
\nonumber\\[1mm]&\qquad
+\kappa^{XS}_{1\,1}\,\kappa^{YS}_{1\,1}
\Big(\kappa^{B}_{2}\,\kappa^{Q}_{2}-\left(\kappa^{BQ}_{1\,1}\right)^{2}\Big)
\nonumber\\[1mm]&\qquad
+\Big(\kappa^{XB}_{1\,1}\,\kappa^{YQ}_{1\,1}+\kappa^{XQ}_{1\,1}\,\kappa^{YB}_{1\,1}\Big)
\Big(\kappa^{BS}_{1\,1}\,\kappa^{QS}_{1\,1}-\kappa^{BQ}_{1\,1}\,\kappa^{S}_{2}\Big)
\nonumber\\[1mm]&\qquad
+\Big(\kappa^{XB}_{1\,1}\,\kappa^{YS}_{1\,1}+\kappa^{XS}_{1\,1}\,\kappa^{YB}_{1\,1}\Big)
\Big(\kappa^{BQ}_{1\,1}\,\kappa^{QS}_{1\,1}-\kappa^{Q}_{2}\,\kappa^{BS}_{1\,1}\Big)
\nonumber\\[1mm]&\qquad
+\Big(\kappa^{XQ}_{1\,1}\,\kappa^{YS}_{1\,1}+\kappa^{XS}_{1\,1}\,\kappa^{YQ}_{1\,1}\Big)
\Big(\kappa^{BQ}_{1\,1}\,\kappa^{BS}_{1\,1}-\kappa^{B}_{2}\,\kappa^{QS}_{1\,1}\Big)
\biggr\}\,.
\end{align}
In each expression, the six terms in curly brackets correspond to the six independent cofactors of the $3\times 3$ matrix~${\bf K}$.
The second- and third-order canonical cumulants in SAM-3.0 mix different conserved charges via cross-cumulants $\kappa(X, B_a)$ contracted with the inverse charge covariance matrix $(K^{-1})^{ab}$.
This contrasts with the SAM-1.0 limit (Sec.~\ref{sec:sam-1}), where the corresponding canonical cumulants depend only on the total-charge cumulants of the accepted charges and not on additional conserved charges [Eqs.~\eqref{eq:sam1-multi-k2}--\eqref{eq:sam1-multi-k3}]; in SAM-1.0 the mixing through $(K^{-1})$ first appears at fourth order.
The SAM-3.0 mixing is generic when the observable couples to different charges with different strengths, as for momentum-space acceptances.

Higher-order canonical cumulants at multiple conserved charges rapidly become too lengthy and thus are not shown explicitly. They can be obtained from Eqs.~(\ref{eq:lambda_linear_system})--(\ref{eq:lambda_recursion_explicit})
and~(\ref{eq:kce-multidim-master}), in particular using the \textsc{Mathematica} implementation~\cite{SAM3code}.

The only model-dependent input required to apply SAM-3.0 is the set of joint grand-canonical cumulants of the chosen acceptance observable(s) with the total conserved charges. All conservation-induced corrections then follow algebraically from Eqs.~(\ref{eq:k2s2-explicit})--(\ref{eq:k11s3-explicit}) and their higher-order generalizations.

\section{Special cases and benchmarks}
\label{sec:applications}

\subsection{Coordinate-space acceptance in a uniform thermal system (SAM-1.0)}
\label{sec:sam-1}

Consider a coordinate-space acceptance defined by a subvolume $V_{\rm in}=\alpha V$ of a spatially uniform thermal system with total volume~$V$, where $\alpha\equiv V_{\rm in}/V$ and $\beta\equiv 1-\alpha$.
SAM-1.0~\cite{Vovchenko:2020tsr,Vovchenko:2020gne} is recovered from the general SAM-3.0 mapping under two assumptions:
(i)~\emph{spatial uniformity}: the cumulant volume densities (susceptibilities) are the same throughout~$V$, so that joint cumulants of extensive quantities within~$V_{\rm in}$ scale as~$\alpha$ relative to those in the full volume;
(ii)~\emph{thermodynamic limit}: the linear dimensions of both $V_{\rm in}$ and $V_{\rm out}=(1-\alpha)V$ are much larger than the correlation length~$\xi$, so that the GCE distributions in the two subsystems factorize and mixed cumulants of quantities from different subsystems vanish at leading order.
These conditions are specific to short-range correlations, such as coordinate-space correlations in thermodynamic limit; for momentum-space cuts the uniformity and the in/out independence generally fail, motivating the full SAM-3.0 treatment.

\subsubsection{Single observable and single conserved charge}
In the one-dimensional setting of Sec.~\ref{sec:formalism} ($d{=}1$, $s{=}1$),
with an arbitrary observable $X$ in the acceptance and a single conserved charge $B_{\rm tot}$,
assumption~(ii) implies $\kappa_{n,m}(X_{\rm in},B_{\rm tot}) = \kappa_{n,m}(X_{\rm in},B_{\rm in})$ for $n\ge 1$, since all cross-correlations with~$B_{\rm out}$ vanish upon expanding $B_{\rm tot}=B_{\rm in}+B_{\rm out}$.
The right-hand side is a pure subsystem cumulant, so assumption~(i) gives $\kappa_{n,m}(X_{\rm in},B_{\rm in}) = \alpha\,\kappa_{n,m}(X_{\rm tot},B_{\rm tot})$.
Combining the two steps:
\eq{\label{eq:sam1-1d-assumpt}
\kappa_{n,m}(X_{\rm in},B_{\rm tot})
= \alpha\,\kappa_{n,m}(X_{\rm tot},B_{\rm tot}), \qquad n\ge 1.
}
When the observable coincides with the accepted conserved charge, $X=B_{\rm in}$ ($B_{\rm tot}=B_{\rm in}+B_{\rm out}$),
assumption~(ii) additionally causes the observable and charge indices to merge, $\kappa_{n,m}(B_{\rm in},B_{\rm in}+B_{\rm out})=\kappa_{n+m}(B_{\rm in})$, since every cross-term involving~$B_{\rm out}$ vanishes by the in/out independence.
Together with~(i) this gives ($n\ge 1$):
\eq{\label{eq:sam-1-assumpt}
\kappa_{n,m}(B_{\rm in},B_{\rm tot})
= \alpha\,\kappa^{B}_{n+m}.
}
Substituting (\ref{eq:sam-1-assumpt}) into Eqs.~\eqref{eq:k1ce}--\eqref{eq:k4ce} yields the SAM-1.0 results of Ref.~\cite{Vovchenko:2020gne}:
\begin{align}
\label{eq:sam1-k1}
\kappa^{B_{\rm in}|B}_1 &= \alpha\,\kappa^{B}_1,\\
\label{eq:sam1-k2}
\kappa^{B_{\rm in}|B}_2 &= \alpha\,\beta\,\kappa^{B}_2,\\
\kappa^{B_{\rm in}|B}_3 &= \alpha\,\beta\,(1-2\alpha)\,\kappa^{B}_3,\\
\kappa^{B_{\rm in}|B}_4 &= \alpha\,\beta\left[(1-3\alpha\beta)\,\kappa^{B}_4
-3\alpha\beta\,\frac{\bigl(\kappa^{B}_3\bigr)^2}{\kappa^{B}_2}\right].
\label{eq:sam1-k4}
\end{align}

\subsubsection{Multiple observables and conserved charges}
In the multidimensional setting of Sec.~\ref{sec:multidim} ($d$ observables, $s$ charges),
the SAM-1.0 assumptions generalize, for $\mi{n}\neq \mi{0}$, to
\eq{\label{eq:sam-1-assumpt-multi}
\kappa^{\bf XB}_{\mi{n}\,\mi{m}}({\bf X}_{\rm in},{\bf B}_{\rm tot})
= \alpha\,\kappa^{\bf XB}_{\mi{n}\,\mi{m}}({\bf X}_{\rm tot},{\bf B}_{\rm tot}).
}
For a single observable $X$,
the second-order diagonal canonical cumulant (\ref{eq:k2-diag}) becomes
\eq{
\kappa^{X|{\bf B}}_2
=\alpha\,\beta\,\kappa^{X_{\rm tot}}_2
+\alpha^{2}\!\left(\kappa^{X_{\rm tot}}_2-\kappa(X_{\rm tot}, B_a)\,(K^{-1})^{ab}\,\kappa(X_{\rm tot}, B_b)\right),
\label{eq:sam1-noncons-k2}
}
decomposing the canonical variance in the acceptance into a subvolume-sampling contribution proportional to $\alpha\beta$
and a global-conservation contribution proportional to $\alpha^{2}$.
As before, the summation over the repeated charge indices $a$ and $b$ is implied.
The expression in the parenthesis is the Schur complement of $\mathbf{K}$~\eqref{eq:Kab} in the
$(s\!+\!1)\times(s\!+\!1)$ joint covariance matrix of $(X_{\rm tot},\mathbf{B}_{\rm tot})$
and equals the canonical variance of $X_{\rm tot}$ in the full system,
reproducing the result of Ref.~\cite{Vovchenko:2020gne}.

For multiple observables, identifying each observable with one of the accepted conserved
charges, $X_i=B^{i}_{\rm in}$ ($i=1,\dots,s$, $d=s$),
Eq.~\eqref{eq:sam-1-assumpt-multi} implies
$\kappa(X_i, B_a)=\alpha\,K_{ia}$
[cf.\ Eq.~\eqref{eq:Kab}], so that the linear-response
coefficient~\eqref{eq:A-def} reduces to
$A_i^{a}=\alpha\,\delta_i^{a}$: the accepted charge is linearly
thinned from the total with fraction~$\alpha$.\footnote{The same
reduction $A_i^{a}=\alpha\,\delta_i^{a}$ arises independently in
the binomial acceptance limit (Sec.~\ref{sec:binomial_limit}), where
it follows from particle-by-particle sampling with
probability~$\alpha$ rather than from spatial uniformity.}
The general canonical cumulants~(\ref{eq:kce1-general})--(\ref{eq:kce3-general}) and (\ref{eq:kce4_general}), with charge indices $i,j,k,l\in\{1,\dots,s\}$, then become
\begin{align}
\label{eq:sam1-multi-k1}
\kappa(B^{i}_{\rm in} | {\bf B}) &= \alpha\,\kappa(B_i),\\[2mm]
\label{eq:sam1-multi-k2}
\kappa(B^{i}_{\rm in}, B^{j}_{\rm in} | {\bf B}) &= \alpha\,\beta\,\kappa(B_i, B_j),\\[2mm]
\label{eq:sam1-multi-k3}
\kappa(B^{i}_{\rm in}, B^{j}_{\rm in}, B^{k}_{\rm in} | {\bf B}) &= \alpha\,\beta\,(1-2\alpha)\,\kappa(B_i, B_j, B_k),\\[2mm]
\label{eq:sam1-multi-k4}
\kappa(B^{i}_{\rm in}, B^{j}_{\rm in}, B^{k}_{\rm in}, B^{l}_{\rm in} | {\bf B}) &= \alpha\,\beta\bigg[
(1-3\alpha\beta)\,\kappa(B_i, B_j, B_k, B_l)
\nonumber\\&
-\alpha\beta\Big(
\kappa(B_i, B_j, B_a)\,(K^{-1})^{ab}\,\kappa(B_k, B_l, B_b)
\nonumber\\&
+\kappa(B_i, B_k, B_a)\,(K^{-1})^{ab}\,\kappa(B_j, B_l, B_b)
\nonumber\\&
+\kappa(B_i, B_l, B_a)\,(K^{-1})^{ab}\,\kappa(B_j, B_k, B_b)
\Big)\bigg],
\end{align}
where $\kappa(B_{i_1},\ldots,B_{i_n})\equiv\kappa^{\bf B}_{\mi{e}_{i_1}+\cdots+\mi{e}_{i_n}}$ denotes the $n$th-order grand-canonical cumulant of the total conserved charges $B^{i_1}_{\rm tot},\ldots,B^{i_n}_{\rm tot}$, and summation over repeated charge indices $a$ and $b$ is implied.

These expressions reproduce SAM-1.0 equations of Ref.~\cite{Vovchenko:2020gne}, originally written in terms of susceptibilities.
Up to third order, the correction factors are identical for all charge components and their correlations.
At fourth order, the three connected-correction terms, corresponding to the last line of Eq.~\eqref{eq:kce4_general} summed over the distinct pairings $(ij|kl)$, $(ik|jl)$, and $(il|jk)$, encode the interplay between global conservation of different charges.
In the single-charge case, Eqs.~\eqref{eq:sam1-multi-k1}--\eqref{eq:sam1-multi-k4} reduce to Eqs.~\eqref{eq:sam1-k1}--\eqref{eq:sam1-k4}.

\subsection{SAM-2.0 limit}
\label{sec:sam-2}

SAM-2.0~\cite{Vovchenko:2021yen} drops the assumption of spatial uniformity and allows one to apply the method to non-uniform systems and momentum-space acceptances.
It retains the assumption that grand-canonical distributions of observables and charges inside and outside the subsystem are uncorrelated, i.e., the total distribution factorizes as a product of the in/out distributions.
In the language of Sec.~\ref{sec:sam-1}, this is the in/out independence~(ii) without the spatial-uniformity assumption~(i), so the $\alpha$-scaling of Eq.~\eqref{eq:sam1-1d-assumpt} no longer applies.

As an example, consider the variance.
Split the conserved charge into its accepted and rejected parts, $B_{\rm tot}=B_{\rm in}+B_{\rm out}$, with the observable $X$ measured inside the acceptance.
The inputs to the SAM-3.0 variance Eq.~\eqref{eq:k2ce} then decompose as $\kappa^{XB}_{1\,1}=\kappa^{XB_{\rm in}}_{1\,1}+\kappa^{XB_{\rm out}}_{1\,1}$ and $\kappa^{B}_2=\kappa^{B_{\rm in}}_2+\kappa^{B_{\rm out}}_2+2\,\kappa^{B_{\rm in}B_{\rm out}}_{1\,1}$, giving
\eq{\label{eq:sam2-k2-split}
\kappa^{X|B}_2=\kappa^{X}_2-\frac{\big(\kappa^{XB_{\rm in}}_{1\,1}+\kappa^{XB_{\rm out}}_{1\,1}\big)^2}{\kappa^{B_{\rm in}}_2+\kappa^{B_{\rm out}}_2+2\,\kappa^{B_{\rm in}B_{\rm out}}_{1\,1}}.
}
The in/out independence sets the cross-covariances $\kappa^{XB_{\rm out}}_{1\,1}$ and $\kappa^{B_{\rm in}B_{\rm out}}_{1\,1}$ to zero, recovering the published SAM-2.0 variances~\cite{Vovchenko:2021yen}
\eq{\label{eq:sam2-k2}
\kappa^{X|B}_2=\kappa^{X}_2-\frac{\big(\kappa^{XB_{\rm in}}_{1\,1}\big)^2}{\kappa^{B_{\rm in}}_2+\kappa^{B_{\rm out}}_2},
\qquad
\kappa^{B_{\rm in}|B}_2=\frac{\kappa^{B_{\rm in}}_2\,\kappa^{B_{\rm out}}_2}{\kappa^{B_{\rm in}}_2+\kappa^{B_{\rm out}}_2},
}
the first for a non-conserved observable such as net protons, the second the accepted-charge case $X=B_{\rm in}$ [where $\kappa^{X}_2=\kappa^{XB_{\rm in}}_{1\,1}=\kappa^{B_{\rm in}}_2$].
Higher orders follow in the same way.
These in/out cross-cumulants are retained in full by SAM-3.0.
Although SAM-2.0 applies to momentum-space acceptances, the in/out independence it assumes does not generally hold there, and SAM-3.0 is then required for an accurate description, as the binomial benchmark of Sec.~\ref{sec:binomial_limit} demonstrates.

\subsection{Binomial acceptance limit}
\label{sec:binomial_limit}

\subsubsection{Single conserved particle number}
A useful analytic benchmark is provided by the binomial acceptance limit, where an observed
conserved charge inside the acceptance is obtained by independent Bernoulli sampling from the corresponding total charge carriers in the full event~\cite{Savchuk:2019xfg}.
In the present setup, we assume that the single charge corresponds to a conserved particle number, i.e. the production of antiparticles is neglected.
Observed particle number corresponds to a subset of the total number, i.e. we take $X \equiv B_{\rm in}$ and assume that, at fixed total charge $B_{\rm tot}$, the conditional distribution of $B_{\rm in}$ is exactly binomial,
\begin{equation}
P(B_{\rm in}\,|\,B_{\rm tot}) = \mathrm{Bin}(B_{\rm tot},B_{\rm in};\alpha)
=\binom{B_{\rm tot}}{B_{\rm in}}\alpha^{B_{\rm in}}(1-\alpha)^{B_{\rm tot}-B_{\rm in}},
\qquad 0\le \alpha\le 1 .
\label{eq:binom}
\end{equation}
In a real system, this situation is realized, for example, in momentum-space acceptances
when momentum-space correlations are negligible~\cite{Kuznietsov:2022pcn,Kuznietsov:2025zvv},
and it provides a stringent check of the SAM-3.0 mapping.
The same Bernoulli thinning underlies the acceptance and efficiency corrections routinely applied to measured cumulants~\cite{Bzdak:2012an,Kitazawa:2016awu} and has been used to interpret measured proton-number fluctuations~\cite{Kitazawa:2011wh,Kitazawa:2012at,Savchuk:2022ljy}.
Note that the binomial distribution~\eqref{eq:binom} requires $B_{\rm tot}\ge 0$, i.e., it applies to a conserved particle number rather than a net charge;
the general case including antiparticles is treated through the multi-species framework of Appendix~\ref{sec:binomial_multi}.

Notably, SAM-2.0 does not correctly reproduce the binomial distribution cumulants given by probability distribution~\eqref{eq:binom} except for the case of Poisson statistics in the reference ensemble~\cite{Vovchenko:2021yen}. 
Here we show that SAM-3.0 resolves this limitation.

Using Eq.~\eqref{eq:binom}, the conditional moment generating function reads
\begin{equation}
\label{eq:bin-mgf-1d}
\Big\langle e^{t B_{\rm in}}\Big\rangle_{B_{\rm tot}}
=\sum_{B_{\rm in}=0}^{B_{\rm tot}} P(B_{\rm in}\,|\,B_{\rm tot})\,e^{tB_{\rm in}}
=\big(1-\alpha+\alpha e^{t}\big)^{B_{\rm tot}}.
\end{equation}
Using Eq.~\eqref{eq:bin-mgf-1d}, the joint grand-canonical CGF can be written in terms of the total-number CGF.
Averaging over $B_{\rm tot}$ in the grand-canonical ensemble,
\begin{align}
\nonumber
G^{\rm bin}_{\rm gce}(t,\lambda)
&\equiv \ln\Big\langle e^{t\,B_{\rm in}+\lambda\,B_{\rm tot}}\Big\rangle_{\rm gce}
= \ln\sum_{B_{\rm tot}=0}^{\infty}P_{\rm gce}(B_{\rm tot})\,
e^{\lambda\,B_{\rm tot}}\!\sum_{B_{\rm in}=0}^{B_{\rm tot}} P(B_{\rm in}|B_{\rm tot})\,e^{t\,B_{\rm in}}
\nonumber\\
&= \ln\sum_{B_{\rm tot}=0}^{\infty}P_{\rm gce}(B_{\rm tot})\,
e^{\lambda\,B_{\rm tot}}\big(1\!-\!\alpha+\alpha\,e^{t}\big)^{B_{\rm tot}}
= \ln\Big\langle e^{\big(\lambda+c(t)\big)\,B_{\rm tot}}\Big\rangle_{\rm gce}
\nonumber\\
&=G_{B_{\rm tot}}\!\big(\lambda+c(t)\big),\label{eq:bin_G_gce}
\end{align}
where
$c(t)\equiv \ln\!\big(1-\alpha+\alpha e^{t}\big)$
is the CGF for binomial thinning of a single carrier
and $G_{B_{\rm tot}}(\lambda)\equiv \ln\big\langle e^{\lambda B_{\rm tot}}\big\rangle_{\rm gce}$ is the grand-canonical CGF of the total conserved number $B_{\rm tot}$.

Since $\lambda$ enters $G^{\rm bin}_{\rm gce}(t,\lambda)=G_{B_{\rm tot}}\!\big(\lambda+c(t)\big)$ only through the combination $u=\lambda+c(t)$, the $\lambda$-derivatives act directly on $G_{B_{\rm tot}}$: $\partial_\lambda^{\,m}G^{\rm bin}_{\rm gce}=\partial_u^{\,m}G_{B_{\rm tot}}(u)\big|_{u=\lambda+c(t)}$.
Setting $\lambda=0$, the remaining $t$-dependence enters solely through $c(t)$ with $c(0)=0$.
Therefore, for $n\ge 1$, applying the Fa\`a di Bruno formula~\eqref{eq:FdB} to the composition with $c(t)$ gives the joint GCE cumulants of $(B_{\rm in},B_{\rm tot})$ in the binomial reference ensemble (taken as bare $\kappa^{B_{\rm in}B}_{n\,m}$ throughout this subsection):
\eq{
\kappa^{B_{\rm in}B}_{n\,m}&\equiv
\partial_t^{\,n}\partial_\lambda^{\,m}G^{\rm bin}_{\rm gce}(t,\lambda)\big|_{t=\lambda=0}
=\left.\partial_t^{\,n}\Big(\partial_u^{\,m}G_{B_{\rm tot}}(u)\Big)\right|_{u=c(t),\,t=0}
\nonumber\\
&=\sum_{r=1}^{n}\kappa^{B}_{m+r}\,B_{n,r}\!\big(c_1,c_2,\ldots,c_{n-r+1}\big)\,,
}
where
$c_k\equiv c^{(k)}(0)$
are the Bernoulli cumulants,
\begin{equation}
c_1=\alpha,\qquad
c_2=\alpha\beta,\qquad
c_3=\alpha\beta(1-2\alpha),\qquad
c_4=\alpha\beta(1-6\alpha\beta),\qquad \beta\equiv 1-\alpha.
\end{equation}
For instance, $\kappa^{B_{\rm in}B}_{0\,m}=\kappa^{B}_m$ and $\kappa^{B_{\rm in}B}_{1\,m}=\alpha\,\kappa^{B}_{m+1}$.
For the remaining diagonal cumulants ($n\ge 2$, $m=0$), up to fourth order one obtains
\begin{align}
\kappa^{B_{\rm in}}_2 &= \alpha^2\kappa^{B}_2+\alpha\beta\,\kappa^{B}_1,
\label{eq:kappa-bin-gce}
\\
\kappa^{B_{\rm in}}_3 &= \alpha^3\kappa^{B}_3+3\alpha^2\beta\,\kappa^{B}_2+c_3\,\kappa^{B}_1,
\\
\kappa^{B_{\rm in}}_4 &=
\alpha^{4}\kappa^{B}_4+6\alpha^{2}c_2\,\kappa^{B}_3+\big(4\alpha c_3+3c_2^{2}\big)\kappa^{B}_2+c_4\,\kappa^{B}_1,
\end{align}
while the mixed cumulants ($m\ge 1$) read
\begin{align}
\kappa^{B_{\rm in}B}_{2\,1} &= \alpha^{2}\kappa^{B}_3 + c_2\,\kappa^{B}_2 ,
\\
\kappa^{B_{\rm in}B}_{3\,1} &= \alpha^{3}\kappa^{B}_4+ 3\alpha c_2\,\kappa^{B}_3+c_3\,\kappa^{B}_2,
\\
\kappa^{B_{\rm in}B}_{2\,2} &= \alpha^{2}\kappa^{B}_4 + c_2\,\kappa^{B}_3 \,.
\label{eq:kappa-bin-gce-last}
\end{align}

In the shorthand of Eq.~\eqref{eq:A-def},
the binomial limit corresponds to a replacement
$A=\kappa^{XB}_{1\,1}/\kappa^{B}_{2}=\alpha$, so that a single number
controls the SAM-3.0 mapping in this limit.
Substituting the binomial-acceptance GCE cumulants~\eqref{eq:kappa-bin-gce}--\eqref{eq:kappa-bin-gce-last} into the SAM-3.0 mapping formulas~\eqref{eq:k1ce}--\eqref{eq:k4ce} yields complete cancellation of all dependence
on $\kappa^{B}_{n\ge 2}$ and reproduces the exact canonical binomial cumulants of $B_{\rm in}$ at fixed $B_{\rm tot}=B_0$:
\begin{align}
\kappa^{B_{\rm in}|B}_1&=\langle B_{\rm in}\rangle=\alpha B_0, \label{eq:binomial-cumulants-1}\\
\kappa^{B_{\rm in}|B}_2&=\beta\,\langle B_{\rm in}\rangle, \\
\kappa^{B_{\rm in}|B}_3&=\beta(1-2\alpha)\,\langle B_{\rm in}\rangle, \\
\kappa^{B_{\rm in}|B}_4&=\beta(1-6\alpha\beta)\,\langle B_{\rm in}\rangle,
\label{eq:binomial-cumulants-4}
\end{align}
and similarly for higher orders.

\subsubsection{Multiple conserved charges}
The multidimensional binomial-acceptance framework, with species-dependent acceptance probabilities and arbitrary observables, is derived in Appendix~\ref{sec:binomial_multi}.
There we obtain the canonical multi-charge binomial cumulants without deriving the constrained multivariate distribution: we compute the unconstrained grand-canonical binomial cumulants $\kappa^{\mathbf{XB}}_{\mi{n}\,\mi{m}}$, for which the thinning factorizes over species, and substitute them into the SAM-3.0 mapping.
This construction extends the multi-species binomial sampling of efficiency-correction formulas~\cite{Bzdak:2012an,Kitazawa:2016awu,Savchuk:2019xfg} to these joint cumulants.
Here we summarize the physical content for universal acceptance and conserved-charge observables.

Consider $s$ conserved charges with universal acceptance probability $\alpha_i\equiv\alpha$, conserved-charge observables $\mathbf{X}=\mathbf{B}_{\rm in}=\sum_i \mathbf{b}_i N^{\rm in}_i$, and charge idempotency $b_{ip}\in\{0,1\}$. These conditions make each accepted component $B_{{\rm in},p}$ an independent Bernoulli sample of the $B_{{\rm tot},p}$ unit-charge carriers, so the exact binomial conditional $P(B_{{\rm in},p}\,|\,B_{{\rm tot},p})= \mathrm{Bin}(B_{{\rm tot},p},B_{{\rm in},p};\alpha)$ holds at fixed $\mathbf{B}_{\rm tot}$. Diagonal canonical cumulants then coincide with the single-charge result~(\ref{eq:binomial-cumulants-1})--(\ref{eq:binomial-cumulants-4}), losing all dependence on the EoS or composition.

The off-diagonal cumulants, which correlate different conserved charges, instead remain sensitive to how charges are distributed among hadron species. This composition information is absent from the total-charge cumulants $\kappa^{B}_n$, $\kappa^{Q}_m$ alone and enters through the joint multiplicity of species carrying both charges, $M^{BQ}_{1\,1}\equiv\sum_i N^{\rm tot}_i\,b_i\,q_i$ (e.g.\ protons contribute $b_p q_p=1$, whereas neutrons and pions contribute zero). Substituting the binomial-acceptance GCE cumulants of Appendix~\ref{sec:binomial_multi} into the SAM-3.0 mapping yields, at leading order,
\begin{equation}
\kappa^{B_{\rm in}Q_{\rm in}|BQ}_{1\,1} = \alpha\beta\,\mean{M^{BQ}_{1\,1}},
\label{eq:s2_ce_k11}
\end{equation}
where $\mean{M^{BQ}_{1\,1}}\equiv\mean{\sum_i N^{\rm tot}_i\,b_i\,q_i}_{\rm gce}$ is the GCE mean of the joint-charge multiplicity. The canonical baryon-charge correlation is therefore proportional to the mean multiplicity of charged baryons (e.g.\ $p$, $\Sigma^+$), irrespective of the underlying equation of state. Higher-order off-diagonal canonical cumulants involve additional joint cumulants of charge-weighted hadron multiplicities probing multi-particle correlations among species carrying more than one conserved charge; explicit formulas, including $\kappa^{B_{\rm in}Q_{\rm in}|BQ}_{2\,1}$ and the corresponding GCE binomial mixed cumulants for $s=2$, are given in Appendix~\ref{sec:binomial_multi}.

\subsection{Microcanonical ensemble: energy conservation in momentum space}
\label{sec:mce}

For momentum-space acceptances, exact energy conservation can be as important as exact charge conservation: the total energy of the event is fixed event by event and is correlated with the number of particles falling inside any momentum cut.
The SAM-3.0 framework handles this case by treating the total energy $E$ as an additional ``conserved charge'' in the same way as the total particle number $N$,\footnote{For the continuous variable $E$, the fixed-value constraint is imposed by a Dirac delta rather than a Kronecker delta; its Fourier representation leads to the same saddle-point structure.} so that the multidimensional formulas of Sec.~\ref{sec:multidim} apply with one observable ($d=1$, the accepted multiplicity $X\equiv N_{\rm in}$, with conjugate source $t_X$) and two exactly conserved quantities ($s=2$, namely $N$ and $E$, with conjugate sources $\lambda_N,\lambda_E$).

To study this construction in a calculable setting, we consider a classical ideal gas of identical nonrelativistic particles of mass $m$ in a volume $V$ at temperature $T$.
The required input for the SAM-3.0 mapping is the joint cumulant generating function of $(N_{\rm in},N,E)$ in the unconstrained (grand-canonical) reference ensemble,
\eq{\label{eq:GE}
G(t_X,\lambda_N,\lambda_E)
= \ln\!\sum_{N=0}^\infty\sum_{N_{\rm in}=0}^{N}\!
\int_0^\infty\!\!dE\;
\Omega(N,E)\,P(N_{\rm in}\,|\,N,E)\;
e^{t_X N_{\rm in}+\lambda_N N+\lambda_E E}\,e^{-E/T},
}
where $\Omega(N,E)$ is the microcanonical density of states for $N$ particles at total kinetic energy $E$, the Boltzmann factor $e^{-E/T}$ provides the GCE weight, and $P(N_{\rm in}\,|\,N,E)$ is the microcanonical conditional probability of finding $N_{\rm in}$ particles inside the longitudinal-momentum acceptance $|p_z|<p_0$ at fixed total $(N,E)$.
Unlike the binomial-acceptance limit of Sec.~\ref{sec:binomial_limit}, energy conservation couples all particles through the shared energy shell $\sum_i p_i^2=2mE$, so $P(N_{\rm in}\mid N,E)$ does not factorize over particles.
The mixed cumulants $\kappa^{XNE}_{n\,j\,k}$ that feed the SAM-3.0 mapping can nevertheless be computed numerically from Eq.~\eqref{eq:GE} by direct differentiation.

For a classical ideal gas of nonrelativistic particles, the density of states reads
\eq{\label{eq:OmegaNE}
\Omega(N,E) = \frac{V^N\,m^{3N/2}}{N!\,(2\pi)^{3N/2}\,\Gamma(3N/2)}\,E^{3N/2-1}\,.
}
The microcanonical conditional probability is the geometric fraction of phase space on the $3N$-dimensional momentum sphere of radius $\sqrt{2mE}$ in which exactly $N_{\rm in}$ of the $N$ longitudinal momenta lie inside $|p_z|<p_0$:
\eq{
P(N_{\rm in}\,|\,N,E)
&= \binom{N}{N_{\rm in}}\frac{\sqrt{2E}m^{-1/2}}{S_{3N}(\sqrt{2mE})}\left[\prod_{i=1}^{N_{\rm in}}\int_{-p_0}^{p_0}\!dp_i\right]
\!\left[\prod_{i=N_{\rm in}+1}^{N}\int_{|p_i|>p_0}\!\!dp_i\right]
\nonumber\\&\quad\times
\int\limits^{\infty}_{-\infty} d^N p_{x}\int\limits^{\infty}_{-\infty} d^N p_{y}\,\delta\left(E - \sum\limits^{3N}_{k=1} \frac{p_k^2}{2m}\right),}
where the coordinate-space contribution cancels out, and $S_{3N}(R) = 2\pi^{3N/2}R^{3N-1}/\Gamma(3N/2)$ is the surface area of the $3N$-dimensional hypersphere.
To evaluate this integral, one can switch to $2N$-dimensional spherical coordinates and integrate over the generalized solid angle and the radial momentum $p_r^2 = \sum_i(p_{x,i}^2+p_{y,i}^2)$ over the whole space.
Since the argument of the $\delta$-function is independent of the angles, one obtains
\eq{\label{eq:Nin_given_NE}
P(N_{\rm in}\,|\,N,E)
&= \binom{N}{N_{\rm in}}\frac{\,\Gamma(3N/2)\,(2mE)^{-N/2}}{(N-1)!\pi^{N/2}}
\left[\prod_{i=1}^{N_{\rm in}}\int_{-p_0}^{p_0}\!dp_i\right]
\!\left[\prod_{i=N_{\rm in}+1}^{N}\int_{|p_i|>p_0}\!\!dp_i\right]
\nonumber\\&\quad\times
\Bigl(1-\sum_{k=1}^N\frac{p_k^2}{2mE}\Bigr)^{\!N-1}\!
\Theta\!\Bigl(2mE-\sum_{k=1}^N p_k^2\Bigr).
}
Here, the density factor and the Heaviside step function come from the filter property of the delta-function on a semi-infinite integration interval.
\begin{figure}
    \centering
    \includegraphics[width=0.98\linewidth]{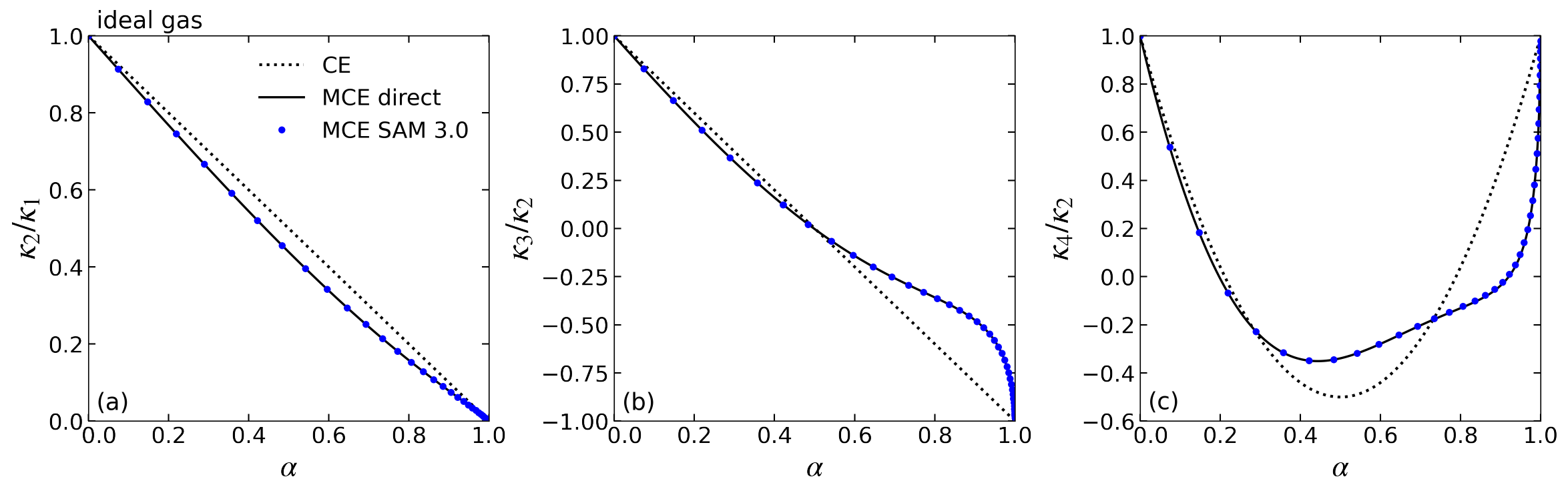}
  \caption{
  Cumulant ratios of the accepted multiplicity $N_{\rm in}$ for a classical ideal gas in the microcanonical ensemble with simultaneous conservation of particle number and energy.
    Panels (a)--(c) show the scaled variance $\kappa_2/\kappa_1$, skewness $\kappa_3/\kappa_2$, and kurtosis $\kappa_4/\kappa_2$ as functions of the accepted particle fraction $\alpha\equiv\langle N_{\rm in}\rangle_{\rm MCE}/N$, set by the longitudinal-momentum cut $|p_z|<p_0$ (with $\alpha=1$ at $p_0=p_{\rm max}$).
    The SAM-3.0 result obtained from Eq.~\eqref{eq:GE} (blue points) is compared with the direct microcanonical large-$N$ prediction of Ref.~\cite{Kuznietsov:2025zvv} (solid black) and with the binomial canonical-ensemble baseline (black dotted), in which only particle number is conserved.
    The SAM-3.0 points lie on top of the direct MCE curve across the full $\alpha$ range.}
    \label{fig:ECONS}
\end{figure}

The conditional probability~\eqref{eq:Nin_given_NE} is evaluated by direct Monte~Carlo sampling on the $3N$-dimensional momentum sphere for each $N$ and $N_{\rm in}\le N$, and is interpolated in $E$ before being inserted into Eq.~\eqref{eq:GE}.\footnote{For an ideal gas of identical classical particles, all dimensionless cumulant ratios are independent of $T$, $V$, and $m$, so we set $T=V=m=1$ throughout the sampling without loss of generality.}
The numerical evaluation underlying Fig.~\ref{fig:ECONS} uses a mean GCE multiplicity $\mean{n}_{\rm gce} = \langle N\rangle_{\rm gce}/V=0.064$, with Monte~Carlo statistics of $5\times10^6$ events per $(N,E)$ bin.
The longitudinal-momentum cut $|p_z|<p_0$ is varied to cover the full acceptance range $0\le\alpha\le 1$.
The joint GCE cumulants $\kappa^{XNE}_{n\,j\,k}\equiv\partial_{t_X}^n\partial_{\lambda_N}^{j}\partial_{\lambda_E}^{k}G\big|_{0}$ are then obtained, and the multidimensional SAM-3.0 mapping of Eqs.~\eqref{eq:k1-diag}--\eqref{eq:k4-diag} (with $d=1$ and $s=2$) yields the microcanonical cumulants of $N_{\rm in}$.

The resulting cumulant ratios are shown in Fig.~\ref{fig:ECONS} and compared with the binomial canonical-ensemble baseline, in which only particle number is conserved.
For the scaled variance, skewness, and kurtosis, the SAM-3.0 result is in quantitative agreement with the large-$N$ analytic prediction of Ref.~\cite{Kuznietsov:2025zvv}, providing a cross-check of the multidimensional SAM-3.0 formalism.
The same construction extends straightforwardly to systems in which total momentum is also conserved, by enlarging the constrained sector to include the three components of the total momentum, i.e. from $(N,E)$ to $(N,E,P_x,P_y,P_z)$. 
The leading second-order correlations between momentum modes induced by exact $(N,E,\mathbf{P})$ conservation in an ideal gas were recently derived in Ref.~\cite{Jaiswal:2026cxt} within a similar conditional saddle-point framework. 
For an ideal gas reference ensemble, the SAM-3.0 mapping reproduces that result at second order [Eq.~\eqref{eq:k2-diag}] and extends it to arbitrary cumulant orders and correlated reference ensembles.
For acceptances that are invariant under ${\bf p}\to-{\bf p}$, the mixed cumulants of the accepted multiplicity with the total momentum vanish by parity, so momentum conservation does not modify the cumulants considered here.

\section{Numerical results}
\label{sec:numerical}

In this section we apply the SAM-3.0 formalism to two concrete calculations:
(i)~updating the hydrodynamics-based non-critical baseline (Hydro-EV) for net-proton cumulants at RHIC-BES, previously obtained with SAM-2.0~\cite{Vovchenko:2021kxx}, and
(ii)~validating the multidimensional SAM-3.0 mapping against direct Monte Carlo sampling with exact conservation of baryon number, electric charge, and strangeness using the Thermal-FIST package~\cite{Vovchenko:2019pjl,Vovchenko:2022hkp}.

\subsection{Hydro-EV baseline with SAM-3.0}
\label{sec:hydroev}

The Hydro-EV baseline~\cite{Vovchenko:2021kxx} provides a non-critical reference for net-proton number cumulants in Au--Au collisions at RHIC-BES energies.
The calculation proceeds in two steps:
(i)~a hydrodynamic simulation with the MUSIC model \cite{Schenke:2010nt,Schenke:2010rr,Paquet:2015lta, Huovinen:2012is}  provides the space-time evolution and the Cooper-Frye particlization hypersurface;
(ii)~the SAM correction maps the resulting grand-canonical proton and antiproton distributions onto canonical (baryon-number-conserving) ones.
In Ref.~\cite{Vovchenko:2021kxx}, step~(ii) was carried out with SAM-2.0, which assumes that the GCE distributions of the accepted and rejected subsystems factorize.

SAM-3.0 removes this assumption.
Instead of separately characterizing the ``in'' and ``out'' subsystems, the formalism of Sec.~\ref{sec:formalism} requires only the joint GCE cumulants $\kappa^{XB}_{n\,m}$ of the observable $X$ (net-proton number in the STAR acceptance) and the total conserved charge $B_{\rm tot}$ (net-baryon number), with a single conserved charge ($s=1$).
The Cooper-Frye formula determines only the mean hadron phase-space density.
Event-by-event fluctuations around this mean are modeled by treating particlization as emission in local grand-canonical equilibrium: the hadron numbers emitted from each hypersurface element fluctuate according to the EV-HRG equation of state at the local temperature and baryochemical potential, with correlations that are short-range in coordinate space~\cite{Vovchenko:2021kxx}.
The Cooper-Frye spectrum of each element fixes the probability for an emitted (anti)proton to fall into the momentum-space acceptance, from which the joint cumulants $\kappa^{XB}_{n\,m}$ follow.
By construction, they incorporate all correlations between the accepted and rejected subsystems, including those arising from momentum-space acceptance cuts.

\begin{figure}[t]
\centering
\includegraphics[width=\textwidth]{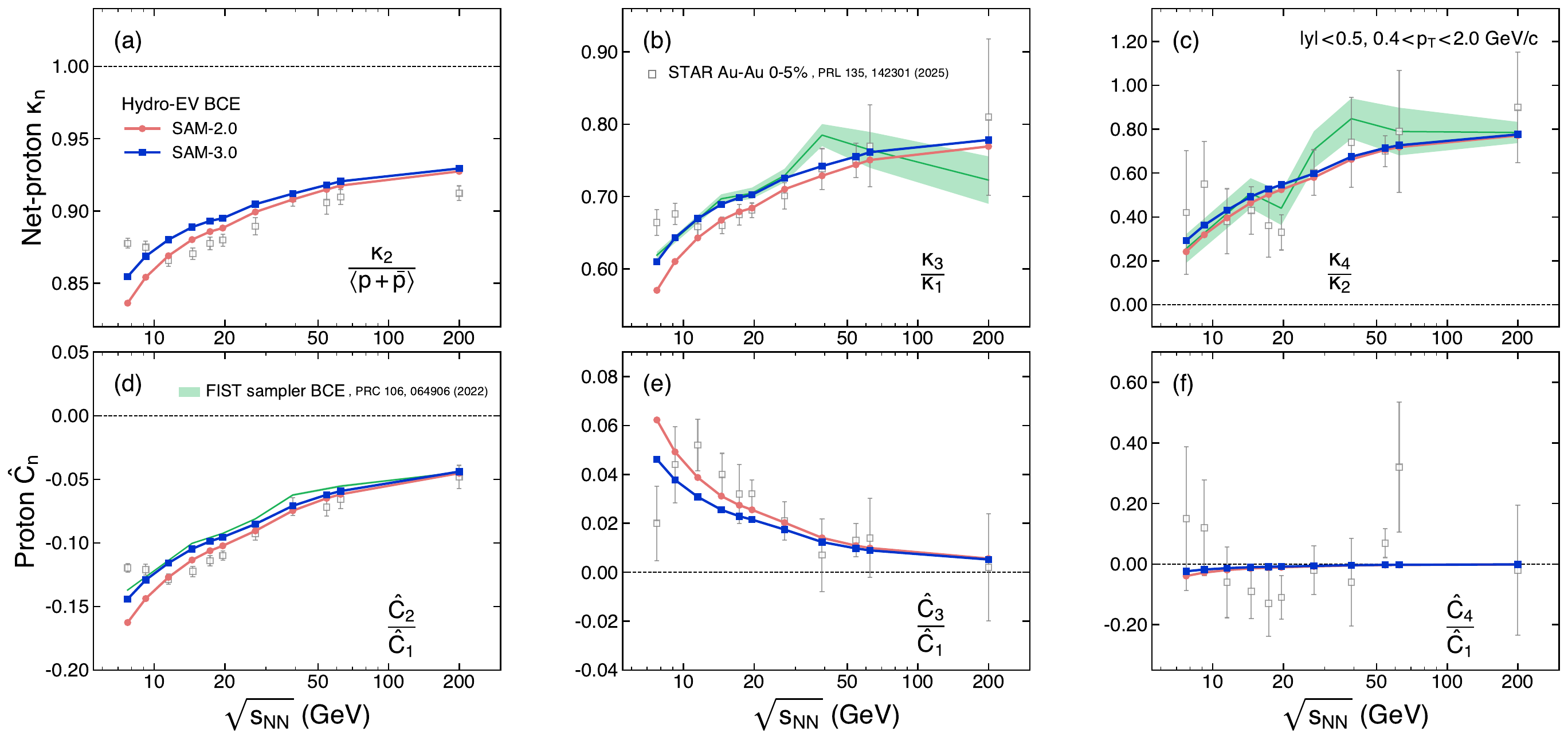}
\caption{
Hydro-EV non-critical baseline for net-proton cumulant ratios (top row) and proton factorial cumulant ratios (bottom row) in central (0--5\%) Au--Au collisions as a function of collision energy $\sNN$.
The SAM-3.0 result with baryon-number conservation (blue squares) is compared with the earlier SAM-2.0 baseline~\cite{Vovchenko:2021kxx} (light red circles), the direct canonical Monte Carlo calculation of Ref.~\cite{Vovchenko:2022hkp} on the same hypersurface using the Thermal-FIST sampler under $B$-only conservation (green band), and STAR data from RHIC-BES-I~($\sNN \geq 39$ GeV)~\cite{STAR:2021iop} and RHIC-BES-II~($\sNN \leq 27$ GeV)~\cite{STAR:2025zdq} (open symbols; the in-figure label refers to the BES-II publication).
The dashed lines in panels~(a) and (d)--(f) indicate the Poisson baseline; in panel~(c) the dashed line marks zero.
The STAR acceptance is $|y|<0.5$, $0.4<p_T<2.0$~GeV/$c$.
}
\label{fig:hydroev}
\end{figure}

We employ the same MUSIC hydrodynamic simulations as in Ref.~\cite{Vovchenko:2021kxx}, covering central Au--Au collisions at $\sNN = 7.7$--$200$~GeV.
As in Ref.~\cite{Vovchenko:2021kxx}, feed-down from resonance decays is included through the decay-inclusive proton-to-baryon fraction in each hypersurface element, with weak-decay contributions counted following the STAR procedure.
The modification of (anti)baryon momentum spectra by decay kinematics is neglected. This approximation was validated in Ref.~\cite{Vovchenko:2021kxx} and is further supported by the agreement with the direct canonical Monte Carlo (Fig.~\ref{fig:hydroev}), in which all decays are performed explicitly.
The STAR acceptance for protons and antiprotons ($|y|<0.5$, $0.4<p_T<2.0$~GeV/$c$) defines the ``in'' subsystem.
The canonical cumulants of net-proton number are then obtained from the iterative SAM-3.0 mapping, Eqs.~(\ref{eq:kappa-ce-general})--(\ref{eq:k4ce}), using the GCE joint cumulants $\kappa^{XB}_{n\,m}$ as input.

The results are shown in Fig.~\ref{fig:hydroev}.
The top row displays net-proton cumulant ratios $\kappa_2/\mean{p+\bar{p}}$, $\kappa_3/\kappa_1$, and $\kappa_4/\kappa_2$ as a function of $\sNN$, while the bottom row shows the corresponding proton factorial cumulant ratios $\hat{C}_n/\hat{C}_1$.
The SAM-3.0 Hydro-EV baseline (blue) is compared with the earlier SAM-2.0 result~\cite{Vovchenko:2021kxx} (red), the direct canonical Monte Carlo calculation of Ref.~\cite{Vovchenko:2022hkp} obtained on the same hypersurface with the Thermal-FIST sampler under $B$-only conservation (green band), and STAR data from RHIC-BES-I~($\sNN \geq 39$ GeV)~\cite{STAR:2021iop} and RHIC-BES-II~($\sNN \leq 27$ GeV)~\cite{STAR:2025zdq} (open symbols).

At high collision energies ($\sNN \gtrsim 39$~GeV), the two baselines are nearly indistinguishable: the effective baryon-number acceptance fraction is small and baryon--antibaryon pair production dominates, making the in/out correlations negligible.
At lower energies, differences emerge.
SAM-3.0 yields cumulant ratios that are systematically closer to the Poisson baseline (dashed lines) than SAM-2.0, indicating that SAM-2.0 slightly overestimated the combined effect of baryon-number conservation and excluded volume on the net-proton cumulants in this regime.
The effect is most pronounced for the third-order ratios $\kappa_3/\kappa_1$ and $\hat{C}_3/\hat{C}_1$ at $\sNN \lesssim 20$~GeV.

The direct canonical Monte Carlo result of Ref.~\cite{Vovchenko:2022hkp}, obtained on the same hypersurface with the Thermal-FIST sampler under $B$-only conservation, is shown as the green band in Fig.~\ref{fig:hydroev}.
It tracks the SAM-3.0 curve to within statistical uncertainty across the full BES range and, wherever the two baselines differ, sits systematically on the SAM-3.0 side, closer to the Poisson baseline than SAM-2.0, most visibly in $\kappa_3/\kappa_1$ and $\hat{C}_3/\hat{C}_1$ at $\sNN \lesssim 20$~GeV.
This pins down the origin of the shift in the results of SAM-2.0 to SAM-3.0: the in/out independence assumption of SAM-2.0 introduces spurious correlations in the momentum-space acceptance, which the canonical MC and SAM-3.0 both avoid.
Adding electric-charge conservation in the same Monte Carlo produces a further modest suppression of (net-)proton cumulants~\cite{Vovchenko:2022hkp} that brings them close to the original SAM-2.0 ($B$-only) baseline of Ref.~\cite{Vovchenko:2021kxx}. 
That is, the missing $Q$-conservation effect in SAM-2.0 partially cancels its $B$-only overestimate.
The earlier Hydro-EV baseline therefore remained accurate in practice, despite SAM-2.0's failure on the binomial-acceptance benchmark of Sec.~\ref{sec:binomial_limit}.

Note that the qualitative SAM-2.0 failure exposed by the binomial benchmark of Sec.~\ref{sec:binomial_limit} only translates into a sizeable numerical correction when the effective acceptance fraction of the conserved charge is large.
For net protons in the STAR acceptance this fraction is small at all BES energies (suppressed by baryon--antibaryon pair production at high $\sNN$ and by the limited rapidity window at low $\sNN$), so the SAM-2.0 and SAM-3.0 baselines remain numerically close, and the original Hydro-EV baseline of Ref.~\cite{Vovchenko:2021kxx} remains a good approximation in this regime.
The SAM-3.0 refinement over SAM-2.0 should become important in regimes that approach the binomial limit, such as wider rapidity acceptances, fixed-target measurements, or identified-particle observables at the LHC.

\subsection{Validation against FIST Monte Carlo}
\label{sec:fist}

As an independent test of the SAM-3.0 formalism introduced in Sec.~\ref{sec:multidim}, we compare analytic SAM-3.0 predictions with direct Monte Carlo sampling in the canonical ensemble using the Thermal-FIST event generator~\cite{Vovchenko:2019pjl,Vovchenko:2022hkp},
for the case of three simultaneously conserved charges (baryon number~$B$, electric charge~$Q$, and strangeness~$S$) and one conserved charge ($B$ only).
This provides a non-trivial test of the SAM-3.0 formalism in the presence of realistic momentum-space acceptance cuts, multiple conserved charges, excluded volume corrections, and hadronic rescattering.

\begin{figure}[t]
\centering
\includegraphics[width=0.95\textwidth]{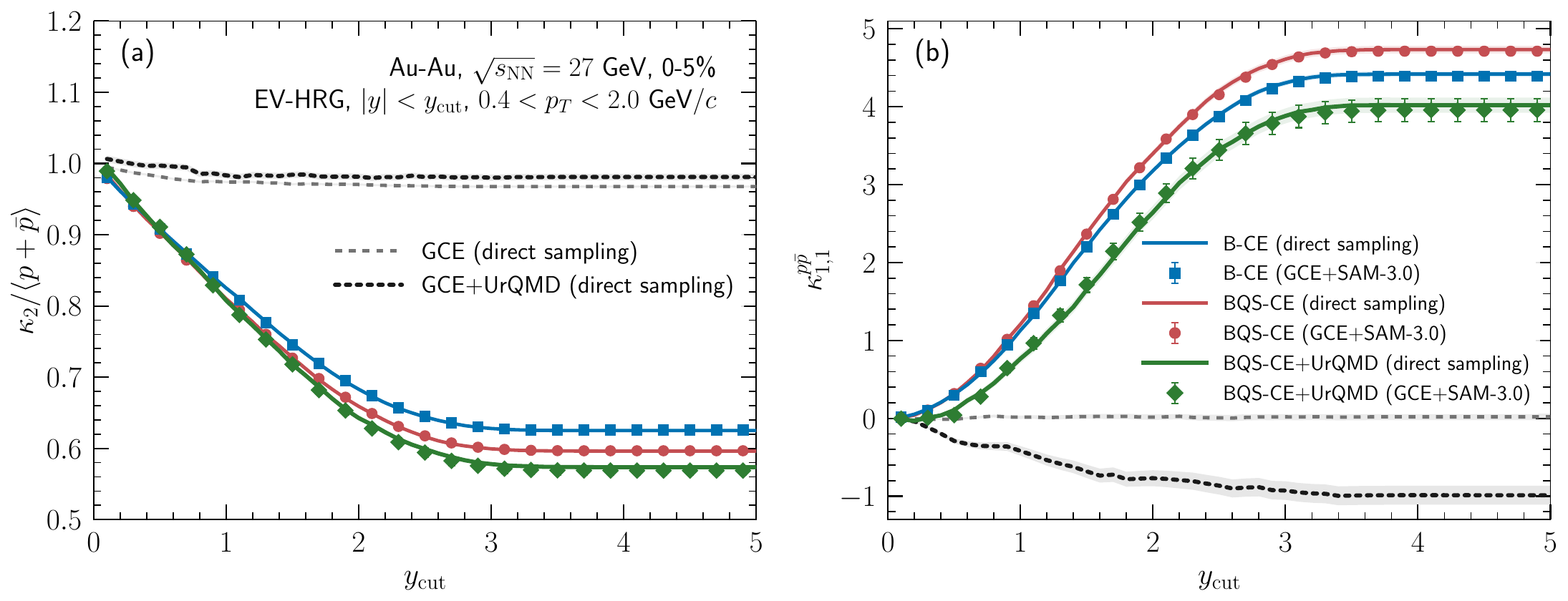}
\caption{Comparison between grand-canonical + SAM-3.0 and direct canonical Monte Carlo sampling on the hypersurface and after hadronic rescattering for central ($0$--$5\%$) Au--Au collisions at $\sNN = 27$~GeV, shown as a function of the rapidity cut $|y|<y_{\rm cut}$. Panel~(a) shows the net-proton scaled variance $\kappa_2/\langle p+\bar p\rangle$, while panel~(b) shows the proton-antiproton joint cumulant $\kappa^{p\bar p}_{1\,1}$.
Lines show direct sampling, markers show grand-canonical sampling plus SAM-3.0.
The grand-canonical EV-HRG baseline is shown without (gray dashed) and with (black dashed) the UrQMD afterburner.
Canonical results are shown for $B$-only conservation (blue) and simultaneous $BQS$ conservation (red), and, after the UrQMD afterburner, for $BQS$ conservation (green); for each, the solid line is direct canonical sampling and the markers are SAM-3.0 ($B$-CE: squares; $BQS$-CE: circles; $BQS$-CE+UrQMD: diamonds).
The SAM-3.0 cumulants use Eqs.~\eqref{eq:k2s3-explicit} and~\eqref{eq:k11s3-explicit} for $BQS$ conservation and Eqs.~\eqref{eq:k2ce} and~\eqref{eq:k11SAMB} for $B$-only conservation. Error bars on the SAM-3.0 points are propagated from the grand-canonical cumulant uncertainties assuming uncorrelated inputs. The transverse-momentum acceptance is $0.4<p_T<2.0$~GeV/$c$.
}
\label{fig:HydroSAM}
\end{figure}

For both these situations we consider two independent Monte Carlo calculations:
\begin{enumerate}
\item \emph{GCE sampling + SAM-3.0.}
Hadrons are sampled from the grand-canonical Cooper-Frye distribution on the hydrodynamic hypersurface. From there, we compute the joint GCE cumulants of the particles both directly after the sampling (including decays) and after a hadronic rescattering phase using UrQMD \cite{Bass:1998ca,Bleicher:1999xi} in the afterburner mode~\cite{urqmd-afterburner-toolkit}. In particular, we compute  $\kappa^{{\bf X}{\bf B}}_{\mi{n}\,\mi{m}}$ with
$\mathbf{X} = (p-\bar{p})$, corresponding to the net-proton observable, and $\mathbf{X} = (p,\bar{p})$, corresponding to the two-dimensional proton-antiproton observable, in the STAR momentum acceptance $0.4<p_T<2.0$~GeV/$c$. The canonical corrections are evaluated for two cases:
simultaneous conservation of $\mathbf{B}=(B_{\rm tot},Q_{\rm tot},S_{\rm tot})$ and conservation of $\mathbf{B}=(B_{\rm tot})$ only. These cumulants are then used as input to the SAM-3.0 mapping of Sec.~\ref{sec:multidim_kappa_ce}.
\item \emph{Canonical Monte Carlo.}
Hadrons are sampled also both directly in the canonical ensemble on the same hypersurface using the \texttt{fist-sampler} of Ref.~\cite{Vovchenko:2022hkp} and after hadronic rescattering. We consider exact conservation of $B$, $Q$, and $S$, as well as exact conservation of $B$ only. The cumulants and joint cumulants are then computed directly from the canonical event sample.
\end{enumerate}
Figure~\ref{fig:HydroSAM} focuses on central collisions ($0$--$5\%$), where the SAM-3.0 formalism is expected to be most accurate, at a representative collision energy of $\sNN = 27$~GeV.
We compare GCE sampling + SAM-3.0 with direct canonical sampling, before and after the afterburner. The SAM-3.0 corrections for three conserved charges are obtained from Eqs.~\eqref{eq:k2s3-explicit} and~\eqref{eq:k11s3-explicit}, and for one conserved charge from Eqs.~\eqref{eq:k2ce} and~\eqref{eq:k11SAMB}.
We show the result as a function of the rapidity acceptance window $|y|<y_{\rm cut}$ for two observables: the net-proton scaled variance $\kappa_2/\langle p+\bar p\rangle$
and the proton-antiproton correlation $\kappa^{p\bar p}_{1\,1}$. The comparison includes simultaneous conservation of $B$, $Q$, and $S$, as well as baryon-number conservation only. 
The grand-canonical baseline, for which the two observables are unaffected by
exact conservation, is shown in Fig.~\ref{fig:HydroSAM} by dashed lines for reference.
The GCE baseline deviates slightly from the Poisson baseline ($\kappa_2/\langle p+\bar p\rangle\simeq 1$ and $\kappa^{p\bar p}_{1\,1}\simeq 0$), a suppression caused by the excluded volume.
For the $p$--$\bar{p}$ correlations, the GCE baseline shows no visible deviation from the Poisson baseline, in which the correlation vanishes.
The excluded volume does not affect the $p$--$\bar{p}$ correlation because the repulsion acts only among like pairs of baryons in this model.

In all cases shown, the exact conservation drives both observables away from the grand-canonical baseline as the acceptance is increased: $\kappa_2/\langle p+\bar p\rangle$ is suppressed from near unity to a plateau of $\sim 0.57$--$0.63$, while the proton-antiproton covariance $\kappa^{p\bar p}_{1\,1}$ increases and saturates at $y_{\rm cut}\gtrsim 3$.
The exact conservation of the three charges (red, circle markers and curves) leads to slightly larger deviations from the GCE baseline than the single conserved charge case (blue, square markers).
This additional deviation reflects global electric-charge conservation, which affects the proton and antiproton numbers.

The hadronic afterburner impacts both the GCE baseline and the CE corrections.
In particular, it leads to a sizable additional anticorrelation between protons and antiprotons relative to the case without the afterburner.
This effect can be explained by proton-antiproton annihilation in the hadronic phase~\cite{Savchuk:2021aog}: it reduces the number of $p$--$\bar{p}$ pairs close enough in phase space, thereby reducing their correlation.
Across the full $y_{\rm cut}$ range, for both observables and cases, the SAM-3.0 corrections reproduce the direct canonical Monte Carlo result within statistical uncertainties, for both $BQS$ and $B$-only conservation.

\begin{figure}[t]
\centering
\includegraphics[width=0.95\textwidth]{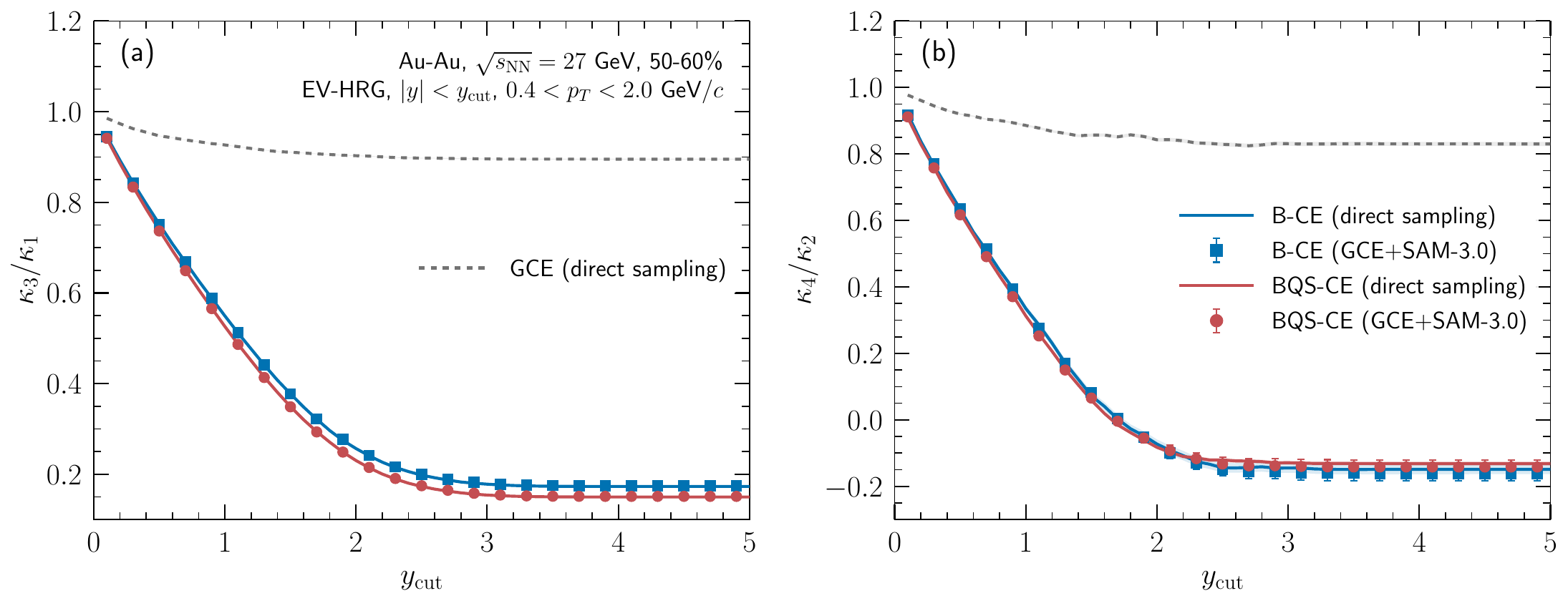}
\caption{
Same as Fig.~\ref{fig:HydroSAM}, but for higher-order net-proton cumulant
ratios in semi-peripheral ($50$--$60\%$) Au--Au collisions at
$\sNN = 27$~GeV.
The left panel shows $\kappa_3/\kappa_1$, while the right panel shows
$\kappa_4/\kappa_2$, both as functions of the rapidity acceptance window
$|y|<y_{\rm cut}$.
Direct canonical Monte Carlo sampling with exact $B$ and with exact $BQS$ conservation is
compared with the corresponding SAM-3.0 results on the same
hypersurface, together with the grand-canonical baseline without exact
conservation effects.
No hadronic afterburner is applied here; other conventions are the same as in Fig.~\ref{fig:HydroSAM}.
}
\label{fig:HydroSAMHigher}
\end{figure}

Figure~\ref{fig:HydroSAMHigher} focuses on non-central collisions ($50$--$60\%$), where the smaller total multiplicity provides a more stringent test of the saddle-point approximation underlying SAM-3.0 than in central collisions, while the multiplicity remains large enough to extract higher-order cumulants reliably. The SAM-3.0 formulas for these ratios are given by Eqs.~\eqref{eq:k3-diag}--\eqref{eq:k4-diag}, with $s=1$ for $B$-only and $s=3$ for $BQS$ conservation; the explicit $s=3$ expressions were generated using the code of Ref.~\cite{SAM3code}.
We compare direct canonical Monte Carlo sampling with exact baryon-number conservation~[$\mathbf{B}=(B_{\rm tot})$] and $BQS$ conservation~[$\mathbf{B}=(B_{\rm tot}, Q_{\rm tot}, S_{\rm tot})$] to GCE sampling + SAM-3.0 for both, focusing on higher-order net-proton cumulant ratios.
For this analysis we do not apply hadronic afterburner. 
The left panel shows $\kappa_3/\kappa_1$, while the right panel shows $\kappa_4/\kappa_2$, both as functions of the rapidity acceptance window. 
The grand-canonical baseline is also shown; in the absence of exact conservation effects, one expects $\kappa_3/\kappa_1 \sim \kappa_4/\kappa_2 \sim 1$. 
The small deviations from unity in the GCE direct sampling reflect the suppression due to the excluded volume. 
As in Fig.~\ref{fig:HydroSAM}, we observe excellent agreement between the two
approaches. 
In particular, the SAM-3.0 formalism for multiple charges also accurately captures the small additional suppression of $\kappa_3/\kappa_1$ due to charge and strangeness conservation.
This demonstrates that SAM-3.0 reproduces canonical higher-order cumulants and remains accurate in the lower-multiplicity regime probed by semi-peripheral collisions.

The excellent agreement validates several aspects of the SAM-3.0 formalism: the saddle-point approximation for one and three conserved charges (Sec.~\ref{sec:multidim}), the iterative recursion for the saddle-point coefficients $\boldsymbol{\lambda}_{\mi{n}}$ [Eqs.~(\ref{eq:lambda_linear_system})--(\ref{eq:lambda_recursion_explicit})], and the master formula for the canonical cumulants [Eq.~(\ref{eq:kce-multidim-master})].
The accurate reproduction of the higher-order cumulants confirms that the method holds across several cumulant orders. Finally, the agreement in the afterburner case demonstrates that the SAM-3.0 method can be used as is to compute canonical corrections in realistic simulation setups with non-equilibrium effects, enabling systematic use in heavy-ion collision studies.

\section{Summary and outlook}
\label{sec:summary}

In this work we introduced the subensemble acceptance method 3.0 (SAM-3.0), which corrects cumulants of an observable measured in a subsystem for the effect of exact global conservation of one or more Abelian charges.
The method takes as input the joint grand-canonical cumulants $\kappa^{XB}_{n\,m}$ of the acceptance observable(s) with the total conserved charge(s), and returns the canonical cumulants at fixed total charge.
It is based on the fact that the canonical ensemble is the grand-canonical ensemble conditioned on a fixed value of the total charge [Eq.~\eqref{eq:Pce_cond}]. 
As a result, the canonical ensemble cumulants can be evaluated, for a large system, through a saddle-point of the joint cumulant generating function $G_{\rm gce}(t,\lambda)$ [Eqs.~\eqref{eq:Gce_integral}, \eqref{eq:Gce_final}].
Because the conservation law enters only through these joint cumulants, the construction requires no separation of the event into accepted and rejected subsystems and places no restriction on their correlation, which makes it applicable to arbitrary momentum-space acceptances.

The corrected cumulants follow from a closed recursion employing uni- and multivariate partial exponential Bell polynomials.
The saddle-point coefficients $\lambda_n$ are obtained from their own recursion [Eqs.~\eqref{eq:lambda_n}, \eqref{eq:lambda_recursion_explicit}] and fed into the cumulant mapping, given by Eqs.~\eqref{eq:kappa-ce-general} and~\eqref{eq:k1ce}--\eqref{eq:k4ce} for a single conserved charge and by the master formula Eq.~\eqref{eq:kce-multidim-master} in general.
The method treats any number of observables, including non-conserved quantities such as net protons, together with any number of simultaneously conserved charges.
The QCD sector $B$, $Q$, $S$ is given explicitly in Eqs.~\eqref{eq:k2s3-explicit} and~\eqref{eq:k11s3-explicit}, and the total energy can be included on the same footing to describe the microcanonical case.

The method contains the earlier SAM frameworks as limiting cases and generalizes them.
For a spatially uniform system with short-range correlations, it reduces to the SAM-1.0~\cite{Vovchenko:2020tsr,Vovchenko:2020gne} formalism describing cumulants in a coordinate-space acceptance [Eqs.~\eqref{eq:sam1-k1}--\eqref{eq:sam1-k4}].
For a more general system where unconstrained cumulants inside and outside acceptance are uncorrelated, the results of SAM-2.0~\cite{Vovchenko:2021yen} are recovered.
SAM-3.0 additionally incorporates correlations in the reference ensemble among observables inside and outside the acceptance.
In particular, the method reproduces the exact canonical cumulants of a uniform binomial acceptance [Eqs.~\eqref{eq:binomial-cumulants-1}--\eqref{eq:binomial-cumulants-4}], 
where SAM-2.0 would produce spurious correlations beyond the binomial distribution.

Although the method has been devised for conserved charges, it also describes microcanonical effects on the cumulants by promoting the total energy to a conserved quantity.
We tested this on the example of a classical ideal gas in the microcanonical ensemble.
The SAM-3.0 formulas applied to the grand-canonical cumulants of particle number and energy were shown to agree with the direct large-$N$ prediction of Ref.~\cite{Kuznietsov:2025zvv} over the full range of the accepted fraction (Fig.~\ref{fig:ECONS}).

We also derived the leading finite-size correction to the saddle-point result.
This correction arises solely from the Gaussian saddle-point prefactor and allows one to verify the accuracy of the leading-order saddle-point result. The result also reproduces the NLO corrections to the cumulants obtained in Ref.~\cite{Barej:2022ccb} in the uniform system limit.

Applying the NLO correction to net-proton cumulants in Au--Au collisions at RHIC-BES, we found that the leading-order results are accurate.
The only possible outliers are antiproton cumulants in peripheral collisions at low energies, where the relative corrections are largest.
This justifies the use of the saddle-point result in the Hydro-EV baseline for proton number cumulants~\cite{Vovchenko:2021kxx}.

We applied the method to two calculations for proton number cumulants in the RHIC Beam Energy Scan.
First, we performed an analytic calculation of an updated non-critical Hydro-EV baseline for net-proton cumulants in central Au--Au collisions at $\sNN = 7.7$--$200$~GeV [Fig.~\ref{fig:hydroev}], taking the joint grand-canonical cumulants on the Cooper-Frye hypersurface as the only input.
The result agrees with direct canonical Monte Carlo sampling~\cite{Vovchenko:2022hkp} within statistical uncertainty throughout the entire energy range.
The refined calculations are slightly closer to the Poisson baseline than the earlier SAM-2.0 calculation~\cite{Vovchenko:2021kxx}, with the difference most pronounced in $\kappa_3/\kappa_1$ and $\hat{C}_3/\hat{C}_1$ below $\sNN \simeq 20$~GeV.
This gap reflects the omission of the in/out correlations in SAM-2.0.

The second application demonstrates the utility of SAM-3.0 within a Monte Carlo event-generator workflow.
We employed \texttt{fist-sampler} to sample Au--Au events at $\sNN = 27$~GeV in the grand-canonical ensemble, optionally followed by an UrQMD afterburner, and used SAM-3.0 to incorporate canonical effects due to global conservation of baryon number, electric charge, and strangeness.
In this workflow, one avoids the numerically costly procedure of enforcing a canonical ensemble at particlization within the event generator.
The SAM-3.0 results reproduce the cumulants obtained from direct sampling in the canonical ensemble.
In particular, SAM-3.0 describes the cumulants accurately once the hadronic phase dynamics are incorporated via UrQMD, which introduces non-equilibrium effects on proton/antiproton cumulants due to baryon annihilation.
This illustrates the generality of the method, namely, that it is not restricted to distributions from equilibrium partition functions in the thermodynamic limit.

One limitation of the current method is its restriction to a global conservation scenario, where conservation laws correlate all particles in the system.
The size of the conservation region in dynamical systems such as heavy-ion collisions can be more restrictive, leading to a concept of local charge conservation~\cite{Bozek:2012en,Sakaida:2014pya,Vovchenko:2024pvk}.
Local charge conservation can potentially be introduced by resolving the conserved-charge density into Fourier modes and constraining each mode separately, with global conservation recovered when only the zero mode (the total charge) is constrained. 
Such an extension can connect SAM-3.0 to the density-correlator formalism of Refs.~\cite{Vovchenko:2024pvk,Ciacco:2026odz}.

The present single-saddle approximation can become unreliable when the standardized higher cumulants are not small, for example in small systems or near phase coexistence, where \(P_{\rm gce}(B_{\rm tot})\) can become bimodal (Sec.~\ref{sec:saddlepoint}). 
The method could be extended to treat a two-phase decomposition of the partition function, where the projected contributions could be evaluated separately and summed, provided each admits a controlled saddle-point approximation. 
When isolated-saddle approximations cease to be valid, an appropriate uniform or other non-Gaussian treatment would be needed. 
We leave such extensions to future work.

The potential applications of SAM-3.0 lie in the calculation of (micro-)canonical effects on cumulants of particle numbers in heavy-ion collisions.
This can be done both on the analytic and Monte Carlo levels.
For instance, SAM-3.0 can be used to avoid costly rejection-sampling steps in Monte Carlo event generators for heavy-ion collisions and calculate the cumulants from grand-canonical simulations only.
The method can also be used to apply canonical corrections in analytical calculations of the cumulants.
An example shown here is the revised Hydro-EV calculation.
Another application is calculating fluctuations of observables from the equation of state, for instance, using the maximum entropy method~\cite{Karthein:2025hvl}.
Until recently, this had been done in equilibrium within the grand-canonical ensemble only~\cite{Karthein:2025hvl,Basar:2026irk}.
Using SAM-3.0 allows one to simultaneously project the effects of the canonical ensemble and the equation of state onto measurable cumulants~(see Ref.~\cite{Pihan:2026xng} for a first application of this kind).
To facilitate such applications, we provide both a supplemental \textsc{Mathematica} notebook and a C++ header-only library, which implement the recursive procedure to compute canonically corrected cumulants of arbitrary observables and conserved charges to the desired arbitrary order~\cite{SAM3code}.

\acknowledgments

This work was supported by the U.S. Department of Energy,
Office of Science, Office of Nuclear Physics, Early Career Research Program under Award Number DE-SC0026065.

\appendix
\section{Finite-size effects}
\label{sec:finite_size}
In Sec.~\ref{sec:formalism} we approximated $\ln I(t)\simeq F_0(t)$ by neglecting the Gaussian prefactor of the saddle-point integration.
Restoring the prefactor produces the next-to-leading-order (NLO) correction to the canonical cumulant generating function.
The expansion parameter is $1/\kappa^B_2$, identified by the saddle width $\delta\lambda=\mathcal O(1/\sqrt{\kappa^B_2})$.
With $F_n(0)=\kappa^B_n$ for $n\ge 2$, the $n$-th Taylor term in Eq.~\eqref{eq:F_Taylor} is of order $\kappa^B_n/(\kappa^B_2)^{n/2}$ at this scale.
The quadratic term ($n=2$) is of order unity and yields the Gaussian.
For $n\ge 3$, single insertions of odd order vanish by symmetry of the Gaussian integration; the leading surviving non-Gaussian contributions to $\ln I$ are products of two cubic terms and a single quartic, both of order $1/\kappa^B_2$.
The Gaussian prefactor is therefore the only contribution at NLO; non-Gaussian effects begin at NNLO.
In an extensive system $\kappa^B_2\propto V$, so the NLO/LO ratio in canonical cumulants is $\mathcal O(1/V)$.

Truncating the saddle expansion of $F(t,\lambda)=G_{\rm gce}(t,\lambda)-\lambda B_0$ at quadratic order in
$\delta\lambda=\lambda-\lambda_\star(t)$ and performing the resulting Gaussian integral along the steepest-descent contour,
one finds
\eq{
  \ln I(t)
  &=
  F_0(t)
  -\frac12\ln(2\pi)
  -\frac12\ln u(t)
  +\mathcal O(1/\kappa^{B}_2),
  \nonumber\\
  u(t)&\equiv F_2(t)=\partial_\lambda^2 F(t,\lambda)\big|_{\lambda=\lambda_\star(t)}
  =\partial_\lambda^2G_{\rm gce}\bigl(t,\lambda(t)\bigr),
  \label{eq:lnI_prefactor}
}
where $u(t)$ is the curvature of $F$ along the steepest-descent contour.
The constant $-\tfrac12\ln(2\pi)$ cancels in $G_{\rm ce}(t)=\ln I(t)-\ln I(0)$, so the canonical CGF splits into
\begin{align}
  G_{\rm ce}(t)
  =
  \underbrace{\Bigl[G_{\rm gce}(t,\lambda(t)) - \lambda(t) B_0\Bigr]}_{G_{\rm ce}^{\rm LO}(t)}
  \;+\;
  \underbrace{\left[-\frac12\ln\!\frac{u(t)}{u(0)}\right]}_{G_{\rm ce}^{\rm NLO}(t)}
  \;+\;\mathcal O(1/\kappa^{B}_2)\,.
  \label{eq:Gce_LO_NLO}
\end{align}
Here and below, we write $\lambda(t)\equiv\lambda_\star(t)$. 
Equations~\eqref{eq:lnI_prefactor} and \eqref{eq:Gce_LO_NLO} generalize the thermodynamic-limit results
$\ln I(t)\simeq F_0(t)$ and $G_{\rm ce}(t)\simeq G_{\rm ce}^{\rm LO}(t)$
[Eq.~\eqref{eq:Gce_final}] to include the leading finite-size correction.

Correspondingly, the canonical cumulants 
$\kappa^{X|B}_n \equiv \partial^n_t G_{\rm ce}(t)\big|_{t=0}$ split as
\eq{
\kappa^{X|B}_n=\kappa^{X|B,{\rm LO}}_n+\kappa^{X|B,{\rm NLO}}_n+\mathcal O(1/\kappa^{B}_2).
}
Denoting the Taylor coefficients of the curvature by $u_n\equiv \partial_t^n u(t)\big|_{t=0}$, such that $u_0=u(0)=\kappa^{B}_{2}$ (since $\lambda(0)=0$),
and applying the Fa\`a di Bruno formula~\eqref{eq:FdB} to $\ln u(t)$, we obtain the compact NLO correction from Eq.~\eqref{eq:Gce_LO_NLO}:
\eq{
  \kappa^{X|B,{\rm NLO}}_n &= \partial^n_t G^{\rm NLO}_{\rm ce}(t)\big|_{t=0}
  = -\frac12\,\frac{d^n}{dt^n}\ln u(t)\Big|_{t=0}
  \nonumber\\
  &=
  -\frac12
  \sum_{k=1}^{n}(-1)^{k-1}(k-1)!\,
  \frac{1}{(\kappa^{B}_{2})^{k}}\,
  B_{n,k}(u_1,\ldots,u_{n-k+1}),
  \quad n\ge1.
  \label{eq:kappa_NLO_Bell}
}
The coefficients $u_n$ for $n\ge1$ are obtained by differentiating $u(t)=\partial_\lambda^2 G_{\rm gce}\!\big(t,\lambda(t)\big)$ using Eq.~\eqref{eq:FdB}:
\begin{equation}
  u_n
  =
  \kappa^{XB}_{n\,2}
  +\sum_{m=0}^{n}\binom{n}{m}\sum_{q=1}^{n-m}
  \kappa^{XB}_{m,q+2}\,
  B_{n-m,q}(\lambda_1,\ldots,\lambda_{n-m-q+1}).
  \label{eq:un_Bell}
\end{equation}
Together, Eqs.~\eqref{eq:lambda_n}, \eqref{eq:kappa-ce-general}, \eqref{eq:kappa_NLO_Bell}, and \eqref{eq:un_Bell} provide an explicit, closed mapping from the mixed GCE cumulants $\kappa^{XB}_{n\,m}$ to the canonical cumulants at a fixed $B=B_0$, accurately capturing the leading finite-size correction from the saddle prefactor.

The first two canonical cumulants, including NLO corrections, read explicitly:
\begin{align}
\label{eq:kappa_1_NLO}
    \kappa^{X|B}_{1} &= \kappa^{X}_{1} - \frac{\kappa^{XB}_{1\,2}}{2\kappa^{B}_{2}} + \frac{\kappa^{B}_{3}\kappa^{XB}_{1\,1}}{2(\kappa^{B}_{2})^2},
    \\\label{eq:kappa_2_NLO}
    \kappa^{X|B}_{2} & =\kappa^{X}_{2} - \frac{(\kappa^{XB}_{1\,1})^2}{\kappa^{B}_{2}} -\frac{1}{2 (\kappa^{B}_{2})^4} \left(
    -2 (\kappa^{B}_{3})^2 (\kappa^{XB}_{1\,1})^2
    + \kappa^{B}_{2} \kappa^{B}_{4} (\kappa^{XB}_{1\,1})^2
    + 4 \kappa^{B}_{2} \kappa^{B}_{3} \kappa^{XB}_{1\,1} \kappa^{XB}_{1\,2}
    \right.\nonumber\\& \quad \left.- (\kappa^{B}_{2})^2 (\kappa^{XB}_{1\,2})^2
    - 2 (\kappa^{B}_{2})^2 \kappa^{XB}_{1\,1} \kappa^{XB}_{1\,3}
    - (\kappa^{B}_{2})^2 \kappa^{B}_{3} \kappa^{XB}_{2\,1}
    + (\kappa^{B}_{2})^3 \kappa^{XB}_{2\,2}
\right).
\end{align}
Higher-order expressions can be generated using the accompanying \textsc{Mathematica} notebook~\cite{SAM3code}.

These expressions simplify considerably for uniform systems with short-range correlations, the core assumptions behind the original SAM-1.0 framework (Sec.~\ref{sec:sam-1}).
NLO corrections for finite system sizes have been worked out in Ref.~\cite{Barej:2022ccb} for this specific case, providing a cross-check of our formalism.
Using Eq.~\eqref{eq:sam-1-assumpt}, one obtains:
\eq{
\label{eq:sam1-k1-prefactor}
    \kappa^{X|B}_1 &= \kappa^{X|B,{\rm LO}}_1,
    \\\kappa^{X|B}_2 &= \kappa^{X|B,{\rm LO}}_2+ \alpha \beta \frac{(\kappa^{B}_{3})^2-\kappa^{B}_{2}\kappa^{B}_{4}}{2(\kappa^{B}_{2})^2},
    \\\kappa^{X|B}_3 &= \kappa^{X|B,{\rm LO}}_3 + \alpha\beta(1-2\alpha)\frac{\kappa^{B}_{3}\kappa^{B}_{4}-\kappa^{B}_{2}\kappa^{B}_{5}}{2(\kappa^{B}_{2})^2},
    \\\nonumber
    \kappa^{X|B}_4 &=  \kappa^{X|B,{\rm LO}}_4 + \frac{\alpha\beta}{2}\left[\frac{\kappa^{B}_{3}\kappa^{B}_{5} - \kappa^{B}_{2}\kappa^{B}_{6}}{(\kappa^{B}_{2})^2}\right.\\
& \quad \left.+ 3\alpha\beta\left(\frac{2(\kappa^{B}_{3})^4 - 5\kappa^{B}_{2}(\kappa^{B}_{3})^2 \kappa^{B}_{4} + (\kappa^{B}_{2})^2 \kappa^{B}_{3}\kappa^{B}_{5}}{(\kappa^{B}_{2})^4} + \frac{(\kappa^{B}_{4})^2 + \kappa^{B}_{2}\kappa^{B}_{6}}{(\kappa^{B}_{2})^2}\right)\right].\label{eq:sam1-k4-prefactor}
}
Here, $\kappa^{X|B,{\rm LO}}_n$ are given by Eqs.~\eqref{eq:k1ce}--\eqref{eq:k4ce}.
Equations~\eqref{eq:sam1-k1-prefactor}--\eqref{eq:sam1-k4-prefactor} correctly reproduce the finite-size NLO corrections of Ref.~\cite{Barej:2022ccb}.

The above NLO formulas generalize straightforwardly to the multidimensional case of $d$ observables and $s$ conserved charges (Sec.~\ref{sec:multidim}).
The one-dimensional curvature $u(t)$ in Eq.~\eqref{eq:lnI_prefactor} is replaced by the $s\times s$ Hessian matrix ${\bf H}(\mathbf{t})\equiv\nabla_{\boldsymbol{\lambda}}^2 G_{\rm gce}(\mathbf{t},\boldsymbol{\lambda})\big|_{\boldsymbol{\lambda}=\boldsymbol{\lambda}(\mathbf{t})}$ along the saddle, with $\mathbf{H}(\mathbf{0})=\mathbf{K}$ recovering the GCE charge cumulant matrix~\eqref{eq:Kab}, yielding\footnote{For real sources and a non-degenerate fluctuating charge basis, $\mathbf{H}(t)$ is positive definite because it is the conserved-charge covariance matrix in the reweighted ensemble. Hence $\det\mathbf{H} > 0$.}
\eq{\label{eq:GNLOmulti}
G^{\rm NLO}_{\rm ce}(\mathbf{t})
= -\frac12\ln\!\left[\frac{\det \mathbf{H}(\mathbf{t})}{\det \mathbf{H}(\mathbf{0})}\right] \,,
}
where the normalization is chosen such that $G_{\rm ce}^{\rm NLO}(\mathbf{0})=0$.
The NLO cumulants $\kappa^{{\bf X}|{\bf B},{\rm NLO}}_{\mi{n}}=\partial_{\mathbf{t}}^{\mi{n}} G^{\rm NLO}_{\rm ce}(\mathbf{t})\big|_{\mathbf{t}=\mathbf{0}}$, with the multi-index $\mi{n}\in\mathbb{N}_0^d$ labeling the cumulant order in each observable direction, follow from differentiating $\ln\det\mathbf{H}(\mathbf{t})$.
Denoting the multivariate Taylor coefficients of $\det\mathbf{H}(\mathbf{t})$ by
$\Delta_{\mi{j}} \equiv \partial^{\mi{j}}_{\mathbf{t}} \det\mathbf{H}(\mathbf{t})\big|_{\mathbf{t} = \mathbf{0}}$
for $\mi{j}\in\mathbb{N}_0^d$ (in particular, $\Delta_{\mi{0}} = \det\mathbf{H}(\mathbf{0})$),
and applying the multivariate Fa\`a di Bruno formula for the composition $\ln\circ\det\mathbf{H}$, one obtains:
\eq{\kappa^{{\bf X}|{\bf B},{\rm NLO}}_{\mi{n}} = -\frac{1}{2}\,\partial^{\mi{n}}_{\mathbf{t}} \left.\ln\det\mathbf{H}(\mathbf{t})\right|_{\mathbf{t} = \mathbf{0}} = -\frac{1}{2}\sum\limits^{|\mi{n}|}_{k=1}\frac{(-1)^{k-1}(k-1)!}{\Delta_{\mi{0}}^{\,k}}\,\mathcal{B}_{\mi{n},k}\!\left(\{\Delta_{\mi{j}}\}\right).}
Here, $\mathcal{B}_{\mi{n},k}$ is the multivariate partial Bell polynomial. Because $\det\mathbf{H}$ is a single scalar function, only one ``color'' enters the colored set partitions of Eq.~\eqref{eq:Bell_multi_explicit}. Thus, the color multi-index $\mi{q}$ reduces to a scalar $k=|\mi{q}|$, and the sum runs over all partitions of the $|\mi{n}|$ derivative directions into exactly $k$ non-empty blocks, with each block $\blk$ contributing a factor $\Delta_{\mi{k}_\blk}$.
The coefficients $\Delta_{\mi{j}}$ can be obtained by applying the multivariate Fa\`a di Bruno formula to $\det\mathbf{H}(\mathbf{t})$ in analogy with Eq.~\eqref{eq:un_Bell}, using the saddle-point derivatives $\boldsymbol{\lambda}_{\mi{k}}$ from the multivariate generalization of Eq.~\eqref{eq:lambda_recursion_explicit}. Explicit results to any order and for any values of $s$ and $d$ can be evaluated using the provided \textsc{Mathematica} notebook~\cite{SAM3code}.

Alternatively, the low-order NLO cumulants can be written in a compact matrix form.
Denoting the Taylor coefficients $\dot{\mathbf{H}}_a\equiv\partial_{t_a}\mathbf{H}(\mathbf{t})\big|_{\mathbf{0}}$, $\ddot{\mathbf{H}}_{ab}\equiv\partial_{t_a}\partial_{t_b}\mathbf{H}(\mathbf{t})\big|_{\mathbf{0}}$,
and utilizing the Jacobi identity $\partial_{t_a}\ln\det\mathbf{H}=\operatorname{Tr}[\mathbf{H}^{-1}\partial_{t_a}\mathbf{H}]$, the NLO correction to the mean reads:
\begin{equation}
\kappa^{{\bf X}|{\bf B},{\rm NLO}}_{\mi{e}_a}
= -\frac{1}{2}\,\operatorname{Tr}\!\left[\mathbf{K}^{-1}\dot{\mathbf{H}}_a\right].
\label{eq:NLO_multi_mean}
\end{equation}
Differentiating once more and substituting $\partial_{t_b}\mathbf{H}^{-1}=-\mathbf{H}^{-1}(\partial_{t_b}\mathbf{H})\mathbf{H}^{-1}$ (obtained from $\mathbf{H}\mathbf{H}^{-1}=\mathbf{I}$), the NLO correction to the (co)variance becomes:
\begin{equation}
\kappa^{{\bf X}|{\bf B},{\rm NLO}}_{\mi{e}_a+\mi{e}_b}
= -\frac{1}{2}\,\operatorname{Tr}\!\left[\mathbf{K}^{-1}\ddot{\mathbf{H}}_{ab}
- \mathbf{K}^{-1}\dot{\mathbf{H}}_a\,\mathbf{K}^{-1}\dot{\mathbf{H}}_b\right],
\label{eq:NLO_multi_var}
\end{equation}
where the cyclic property of the trace has been used to symmetrize the second term.
The matrix elements of $\dot{\mathbf{H}}_a$ and $\ddot{\mathbf{H}}_{ab}$ follow from the multivariate chain rule applied to $H_{pq}(\mathbf{t})=\partial_{\lambda_p}\partial_{\lambda_q}G_{\rm gce}(\mathbf{t},\boldsymbol{\lambda}(\mathbf{t}))$, evaluated using the saddle-point derivatives $\boldsymbol{\lambda}_{\mi{k}}$.

\section{NLO finite-size estimates for Hydro-EV}
\label{sec:nlo_numerical}

The NLO corrections derived in Appendix~\ref{sec:finite_size} provide an estimate of the leading finite-size error in the SAM-3.0 mapping.
We evaluate them numerically for the Hydro-EV setup of Sec.~\ref{sec:hydroev}.

For baryon-number conservation ($s=1$), the NLO corrections to the first two canonical cumulants are given by Eqs.~(\ref{eq:kappa_1_NLO})--(\ref{eq:kappa_2_NLO}), while the higher orders follow from Eqs.~(\ref{eq:kappa_NLO_Bell}) and~(\ref{eq:un_Bell})~(see also the accompanying \textsc{Mathematica} notebook~\cite{SAM3code}).
The correction is controlled by the GCE baryon number variance $u_0 = \kappa^{B}_2$ and its $t$-derivatives $u_n$.
Since $\kappa^{B}_2$ scales with the system volume, the NLO correction to the canonical cumulants is $\mathcal{O}(1/V)$ in relative terms, so it is largest for the smallest fireball size. 
This occurs in the most peripheral collisions at the lowest accessible energy, which is $\sNN = 7.7$~GeV in the RHIC-BES program in collider mode.

\begin{figure}[t]
\centering
\includegraphics[width=\textwidth]{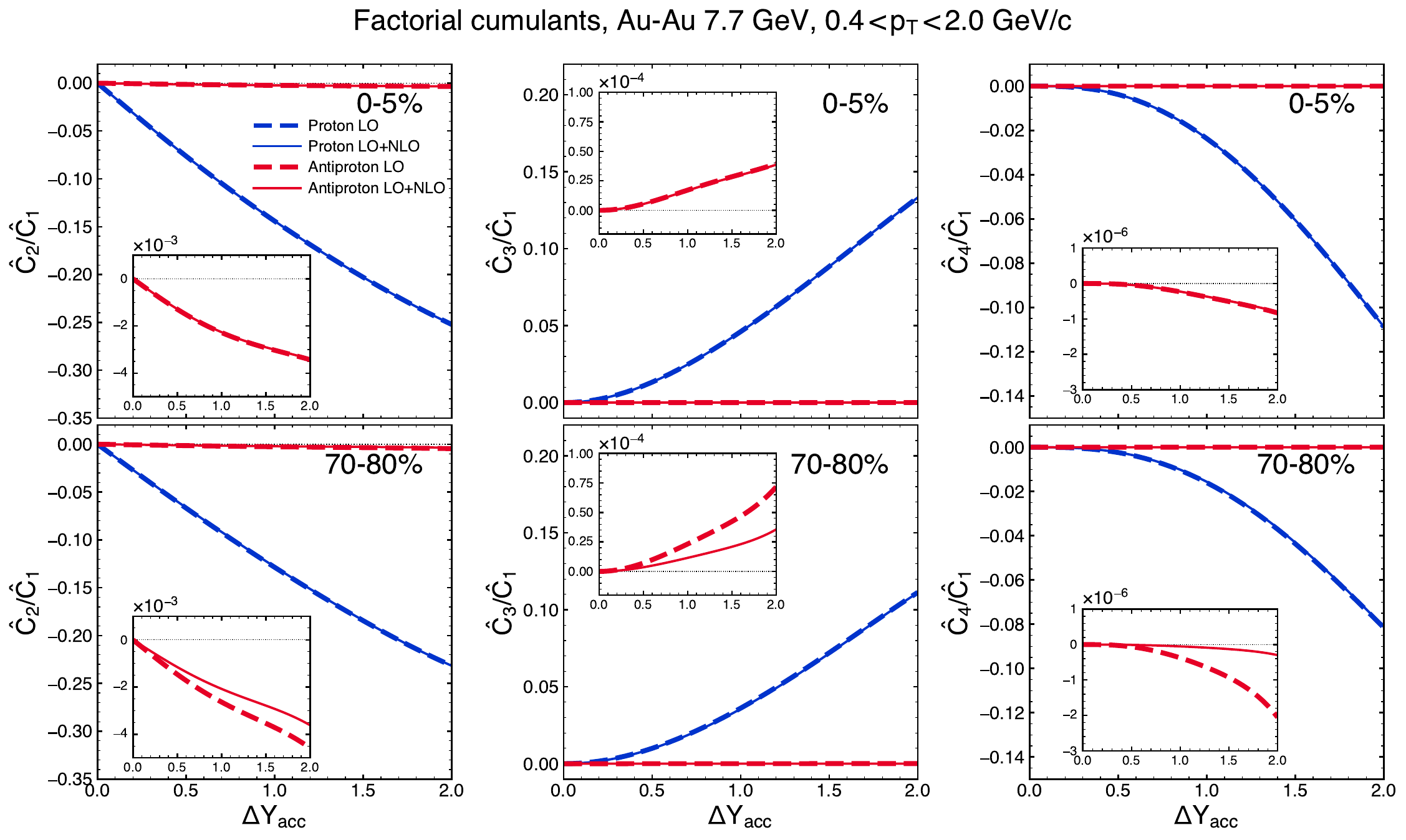}
\caption{
Hydro-EV calculations of the proton (blue) and antiproton (red) factorial cumulant ratios $\hat{C}_n/\hat{C}_1$ for $n=2,3,4$, evaluated on the MUSIC particlization hypersurface, versus the rapidity acceptance window $\Delta Y_{\rm acc}$ (centered at midrapidity, $0.4<p_T<2.0$~GeV/$c$) for Au--Au collisions at $\sNN=7.7$~GeV using SAM-3.0.
Top row: 0--5\% centrality. Bottom row: 70--80\% centrality.
Dashed lines: leading-order SAM-3.0. Solid lines: leading order plus the NLO finite-size correction of Appendix~\ref{sec:finite_size}.
The insets magnify the antiproton curves, which are suppressed in magnitude by the small antiproton abundance at this energy.
}
\label{fig:nlo_7p7}
\end{figure}

We compare two centrality classes at $\sNN = 7.7$~GeV: the central 0--5\% bin used for net-proton cumulant measurements at BES, and the peripheral 70--80\% bin, where the fireball is smallest and the NLO/LO ratio is largest.
Figure~\ref{fig:nlo_7p7} shows the factorial cumulant ratios $\hat{C}_n/\hat{C}_1$ for protons and antiprotons separately, plotted against the rapidity acceptance window $\Delta Y_{\rm acc}$ centered at midrapidity (with $0.4<p_T<2.0$~GeV/$c$ fixed); the top row is 0--5\%, the bottom row 70--80\%.
Dashed lines are the leading-order result while the solid lines add the NLO finite-size correction described in Appendix~\ref{sec:finite_size}.
Factorial cumulants are sensitive to both the excluded-volume correlations carried by the GCE input and to global baryon-number conservation; for this acceptance the latter dominates. The gap between the two curves measures the saddle-point prefactor contribution to the conservation correction; the EV part of the GCE input is the same in LO and LO+NLO.

For protons, the LO and LO+NLO curves overlap to within the line width in both centralities and across the full acceptance range: the relative NLO correction stays well below the percent level.
For antiprotons, the absolute values of $\hat{C}_n/\hat{C}_1$ at $\sNN = 7.7$~GeV are themselves small: of order $\mathcal{O}(10^{-3})$ for $n=2$, $\mathcal{O}(10^{-5})$ for $n=3$, and $\mathcal{O}(10^{-6})$ for $n=4$.
In the central 0--5\% bin the antiproton LO and LO+NLO curves overlap in the insets as well.
In peripheral 70--80\% events the antiproton NLO shift becomes visible there, as expected from the smaller $\kappa^{B}_2$, but it stays negligible in absolute terms.
Since the protons dominate net-proton observables, the conclusion is the same in both bins, and at any other point in the BES program $\kappa^{B}_2$ is larger still.

We note, however, that other observables may have larger NLO corrections.
For example, rare particle species with small GCE variances (e.g., net kaons or net lambdas affected by strangeness conservation) can receive larger corrections, and the formulas of Appendix~\ref{sec:finite_size} can be used to evaluate those as needed.

\section{Binomial acceptance limit (multidimensional case)}
\label{sec:binomial_multi}

We derive the joint cumulants of accepted observables with the full-event conserved charges in the binomial-acceptance limit used in Sec.~\ref{sec:binomial_limit}.

Consider hadron species $i=1,\dots,M$ with full-phase-space multiplicities $N_i^{\rm tot}$.
Each species carries a vector of conserved charges
\begin{equation}
\label{eq:q-n}
\mathbf{b}_i = \big(b_{i1},\dots,b_{is}\big),
\qquad
\mathbf{B}_{\rm tot} \equiv \sum_{i=1}^M \mathbf{b}_i\,N^{\rm tot}_i .
\end{equation}
Then an observable vector $\mathbf{X}\in\mathbb{Z}^d$ is constructed
from accepted multiplicities $N^{\rm in}_i$ as
\begin{equation}
\label{eq:x-n}
\mathbf{X} \equiv \sum_{i=1}^M \mathbf{x}_i\, N^{\rm in}_i ,
\qquad
\mathbf{x}_i\in\mathbb{Z}^d,
\end{equation}
where $\mathbf{x}_i$ encodes how species $i$ contributes to the chosen set of observables
(e.g., proton multiplicity, net-proton, identified-particle yields, etc.).

Inside the experimental acceptance, the observed multiplicities $N^{\rm in}_i$ are obtained by independent
thinning of each species,
\begin{equation} \label{eq:binomial}
P(N^{\rm in}_i \mid N^{\rm tot}_i) = \mathrm{Bin}(N^{\rm tot}_i, N^{\rm in}_i;\alpha_i)=\binom{N^{\rm tot}_i}{N^{\rm in}_i}\alpha_i^{N^{\rm in}_i}(1-\alpha_i)^{N^{\rm tot}_i-N^{\rm in}_i}, \qquad i=1,\dots,M,
\end{equation}
with species-dependent single-particle acceptance probabilities $\alpha_i\in[0,1]$.
In addition, we assume that the acceptance decisions are independent across different species.
Conditioned on $\{N^{\rm tot}_i\}$, the random variables
$\{N^{\rm in}_i\}$ factorize,
\begin{equation}
\label{eq:factorization}
P(\{N^{\rm in}_i\}\mid \{N^{\rm tot}_i\}) = \prod_{i=1}^M P(N^{\rm in}_i\mid N^{\rm tot}_i),
\end{equation}
which is a defining property of the binomial acceptance limit.

Introducing source vectors $\mathbf{t}\in\mathbb{R}^d$ and $\boldsymbol{\lambda}\in\mathbb{R}^s$ conjugate to $\mathbf{X}$ and
$\mathbf{B}_{\rm tot}$, respectively, we define the joint CGF as
\begin{equation}
\label{eq:bin-cgf-generic}
G_{\mathbf{X},\mathbf{B}_{\rm tot}}(\mathbf{t},\boldsymbol{\lambda})
\equiv
\ln\Big\langle e^{\mathbf{t}\cdot\mathbf{X} + \boldsymbol{\lambda}\cdot\mathbf{B}_{\rm tot}} \Big\rangle .
\end{equation}

Conditioning on $\{N^{\rm tot}_i\}$ and using Eqs.~(\ref{eq:q-n}) and~(\ref{eq:x-n}) together with the conditional independence assumption~(\ref{eq:factorization}) yields
\begin{align}
\label{eq:bin-mgf}
\Big\langle e^{\mathbf{t}\cdot\mathbf{X} + \boldsymbol{\lambda}\cdot\mathbf{B}_{\rm tot}} \Big\rangle
&=
\Big\langle \Big\langle e^{\mathbf{t}\cdot \sum_i \mathbf{x}_i N^{\rm in}_i + \boldsymbol{\lambda}\cdot \sum_i \mathbf{b}_i N^{\rm tot}_i}
\Big\rangle_{\{N^{\rm in}_i\}\mid\{N^{\rm tot}_i\}} \Big\rangle_{\{N^{\rm tot}_i\}} \nonumber\\
&=
\Big\langle e^{\sum_i (\boldsymbol{\lambda}\cdot\mathbf{b}_i) N^{\rm tot}_i}
\prod_{i=1}^M \Big\langle e^{(\mathbf{t}\cdot\mathbf{x}_i) N^{\rm in}_i}\Big\rangle_{N^{\rm in}_i\mid N^{\rm tot}_i}
\Big\rangle_{\{N^{\rm tot}_i\}} .
\end{align}
Substituting Eq.~(\ref{eq:binomial}), we obtain
\begin{align}
\label{eq:bin-mgf-x}
\Big\langle e^{(\mathbf{t}\cdot\mathbf{x}_i) N^{\rm in}_i}\Big\rangle_{N^{\rm in}_i\mid N^{\rm tot}_i}
&= \sum_{N^{\rm in}_i=0}^{N^{\rm tot}_i}\binom{N^{\rm tot}_i}{N^{\rm in}_i}\alpha_i^{N^{\rm in}_i}(1-\alpha_i)^{N^{\rm tot}_i-N^{\rm in}_i}e^{(\mathbf{t}\cdot\mathbf{x}_i)N^{\rm in}_i}
\nonumber\\
&= \big(1-\alpha_i+\alpha_i e^{\mathbf{t}\cdot\mathbf{x}_i}\big)^{N^{\rm tot}_i}.
\end{align}
Using Eqs.~(\ref{eq:bin-mgf}) and~(\ref{eq:bin-mgf-x}), the CGF~(\ref{eq:bin-cgf-generic}) can be written as
\begin{equation}
G_{\mathbf{X},\mathbf{B}_{\rm tot}}(\mathbf{t},\boldsymbol{\lambda})
=
\ln\Big\langle \exp\Big(\sum_{i=1}^M \theta_i(\mathbf{t},\boldsymbol{\lambda})\,N^{\rm tot}_i\Big)\Big\rangle_{\{N^{\rm tot}_i\}} ,
\end{equation}
with effective sources
\begin{equation}
\label{eq:theta}
\theta_i(\mathbf{t},\boldsymbol{\lambda})
\equiv
\boldsymbol{\lambda}\!\cdot\!\mathbf{b}_i
+
\ln\!\big(1-\alpha_i+\alpha_i e^{\mathbf{t}\cdot\mathbf{x}_i}\big).
\end{equation}

Equivalently, introducing the CGF of the full multiplicity vector $\mathbf{N}=(N^{\rm tot}_1,\dots,N^{\rm tot}_M)$,
\begin{equation}
G_{\mathbf{N}}(\boldsymbol{\theta})\equiv \ln\Big\langle \exp\Big(\sum_{i=1}^M \theta_i N^{\rm tot}_i\Big)\Big\rangle_{\{N^{\rm tot}_i\}},
\end{equation}
one can write $G_{\mathbf{X},\mathbf{B}_{\rm tot}}(\mathbf{t},\boldsymbol{\lambda})=G_{\mathbf{N}}\!\big(\boldsymbol{\theta}(\mathbf{t},\boldsymbol{\lambda})\big)$, $\boldsymbol{\theta}\in\mathbb{R}^M$.

The joint grand-canonical (reference) cumulants of $(\mathbf{X},\mathbf{B}_{\rm tot})$ are defined as mixed derivatives
of the CGF,
\begin{equation}
\kappa^{\bf XB}_{\mi{n}\,\mi{m}}
\equiv
\left.
\partial_{\mathbf{t}}^{\mi{n}}\,
\partial_{\boldsymbol{\lambda}}^{\mi{m}}\,
G_{\mathbf{X},\mathbf{B}_{\rm tot}}(\mathbf{t},\boldsymbol{\lambda})
\right|_{\mathbf{t}=\boldsymbol{\lambda}=\mathbf{0}},
\qquad
\mi{n}\in\mathbb{N}_0^d,\ \mi{m}\in\mathbb{N}_0^s ,
\label{eq:kappa_gce_def_multi}
\end{equation}
where, using the multiindex notation,
$\partial_{\mathbf{t}}^{\mi{n}}\equiv \prod_{a=1}^d \partial_{t_a}^{n_a}$,
$\partial_{\boldsymbol{\lambda}}^{\mi{m}}\equiv \prod_{b=1}^s \partial_{\lambda_b}^{m_b}$.

To obtain explicit expressions in the binomial acceptance model, we use the composition
$G_{\mathbf{X},\mathbf{B}_{\rm tot}}(\mathbf{t},\boldsymbol{\lambda})
=G_{\mathbf{N}}(\boldsymbol{\theta}(\mathbf{t},\boldsymbol{\lambda}))$.
Introduce the species cumulants of the full-phase-space multiplicity vector $\mathbf{N}=(N^{\rm tot}_1,\dots,N^{\rm tot}_M)$ as
\begin{equation}
\kappa^{\mathbf{N}}_{\mi{n}}
\equiv
\left.
\partial_{\boldsymbol{\theta}}^\mi{n} G_{\mathbf{N}}(\boldsymbol{\theta})
\right|_{\boldsymbol{\theta}=\mathbf{0}} .
\label{eq:kappaN_def}
\end{equation}
All mixed cumulants $\kappa^{\bf XB}_{\mi{n}\,\mi{m}}$ follow from the multivariate chain rule and set-partition combinatorics.

Using Eq.~(\ref{eq:theta}) one has at $\mathbf{t}=\boldsymbol{\lambda}=\mathbf{0}$:
\begin{align}
&\left.\frac{\partial \theta_i(\mathbf{t},\boldsymbol{\lambda})}{\partial \lambda_b}\right|_0 = b_{ib}, \quad
\left.\frac{\partial^m \theta_i(\mathbf{t},\boldsymbol{\lambda})}{\partial \lambda_{b_1}\cdots\partial \lambda_{b_m}}\right|_0=0~~(m\ge 2),
\nonumber\\
&\left.\frac{\partial^n \theta_i(\mathbf{t},\boldsymbol{\lambda})}{\partial t_{a_1}\cdots\partial t_{a_n}}\right|_0\equiv\mathcal{V}^{(n)}_{i;\,a_1\cdots a_n}
=c_n(\alpha_i)\,x_{i a_1}\cdots x_{i a_n}~~(n\ge 1),
\label{eq:theta_derivs}
\end{align}
and all mixed $t$--$\lambda$ derivatives vanish, e.g., $\left.\partial_{t_a}\partial_{\lambda_b}\theta_i\right|_0=0$.
Here $\mathcal{V}^{(n)}_{i;\,a_1\cdots a_n}$ is the rank-$n$ acceptance tensor of species $i$, with
\begin{equation}
c_n(\alpha)\equiv \left.\frac{d^n}{dz^n}\ln\!\big(1-\alpha+\alpha e^{z}\big)\right|_{z=0}\,.
\label{eq:cn_def_all}
\end{equation}
Explicitly,
\begin{align}
c_1(\alpha)=\alpha, \quad
c_2(\alpha)=\alpha \beta, \quad
c_3(\alpha)=\alpha\beta(1-2\alpha), \quad
c_4(\alpha)=\alpha\beta\big[1-6\alpha\beta\big], \ \ldots
\label{eq:cn_examples}
\end{align}

Because of Eq.~\eqref{eq:theta_derivs}, blocks cannot mix $t$- and $\lambda$-derivatives acting on the same $\theta_i$.
Let $\Pi_n$ denote the set of all set partitions of $\{1,\dots,n\}$.
A partition $\pi\in\Pi_n$ is a collection of disjoint nonempty blocks
$\blk\in\pi$ whose union is $\{1,\dots,n\}$.
We denote by $|\pi|$ the number of blocks in the partition $\pi$, and by $|\blk|$ the size
(cardinality) of a block $\blk$.
For a block $\blk=\{j_1,\dots,j_{|\blk|}\}$ we use the shorthands $a_\blk\equiv (a_j)_{j\in \blk}$ and
$\mathcal{V}^{(|\blk|)}_{i_{\blk};\,a_{\blk}}
\;\equiv\;
\mathcal{V}^{(|\blk|)}_{i_{\blk};\,a_{j_1}\cdots a_{j_{|\blk|}}}.$
With this notation,
the general mixed cumulants read

\eq{
\kappa^{\bf XB}_{\mi{n}\,\mi{m}}&=\kappa\!\big(X_{a_1},\dots,X_{a_n};B_{p_1},\dots,B_{p_m}\big)
\nonumber\\
&=
\sum_{\pi\in\Pi_n}
\ \sum_{\substack{\,i_\blk\ (\blk\in\pi)\\ j_1,\dots,j_m}}
\kappa^{\mathbf N}_{\,\{i_\blk\}_{\blk\in\pi},\,j_1,\dots,j_m}\;
\Bigg[\prod_{\blk\in\pi}\mathcal V^{(|\blk|)}_{i_\blk;\,a_\blk}\Bigg]
\Bigg[\prod_{r=1}^m b_{j_r p_r}\Bigg].
\label{eq:binomial-master}
}

The general formula~\eqref{eq:binomial-master} applies to any integer-valued $\mathbf{b}_i$, including non-idempotent net charges; the closed-form expressions below are the idempotent specialization, in which the accepted charge becomes \emph{exactly} binomial at fixed total charge and the resulting compactness is the basis of the binomial benchmark of Sec.~\ref{sec:binomial_limit}. We now specialize Eq.~\eqref{eq:binomial-master} to the physically important case of universal acceptance ($\alpha_i\equiv\alpha$), conserved-charge observables ($\mathbf{X}=\mathbf{B}_{\rm in}$), and charge idempotency ($b_{ip}\in\{0,1\}$). For any multiindex $\mi{u}\in\mathbb{N}_0^s\setminus\{\mi{0}\}$ we define the monomial-weighted multiplicity sum
\begin{equation}
M_{\mi{u}}
\equiv
\sum_{i=1}^M N^{\rm tot}_i \prod_{r=1}^s b_{ir}^{\,u_r}\,,
\label{eq:Mu_def_app}
\end{equation}
which counts hadron multiplicities weighted by monomials of their charge quantum numbers and probes how different conserved charges are co-localized on the same species. For $s=2$ charges $\mathbf{B}=(B,Q)$, $\mathbf{b}=(b,q)$, we adopt the charge-label shorthand $M^{BQ}_{nm}\equiv M_{\mi{u}=(n,m)}$ (same convention as for cumulants, cf.\ Sec.~\ref{sec:multidim_setup}), so that
\begin{equation}
M^{B}_{1}=B_{\rm tot},\qquad M^{Q}_{1}=Q_{\rm tot},\qquad M^{BQ}_{1\,1}=\sum_i N^{\rm tot}_i\,b_i\,q_i\,.
\label{eq:Mu_examples_app}
\end{equation}
We denote joint GCE cumulants of monomial-weighted multiplicities and conserved charges by listing their arguments explicitly: $\kappa(M^{BQ}_{1\,1})$, $\kappa(M^{BQ}_{1\,1},B_{\rm tot})$, $\kappa(M^{BQ}_{1\,1},B_{\rm tot},B_{\rm tot})$, and so on. In practice, the $\kappa(M,\dots)$ can be obtained from any microscopic or thermal model providing event-by-event species multiplicities (hadronic transport, afterburners, event generators, HRG).

For $s=2$ charges, Eq.~\eqref{eq:binomial-master} yields the off-diagonal GCE binomial mixed cumulants
\begin{align}
\kappa^{B_{\rm in}Q_{\rm in}}_{1\,1}
&=
\alpha\Big[
\alpha\,\kappa^{BQ}_{1\,1}
+\beta\,\kappa(M^{BQ}_{1\,1})
\Big],
\label{eq:s2_M_k11}
\\[1mm]
\kappa^{B_{\rm in}Q_{\rm in}}_{2\,1}
&=
\alpha\Big[
\alpha^2\,\kappa^{BQ}_{2\,1}
+\alpha\beta\,\kappa^{BQ}_{1\,1}
+2\alpha\beta\,\kappa(M^{BQ}_{1\,1},B_{\rm tot})
+\beta(1-2\alpha)\,\kappa(M^{BQ}_{2\,1})
\Big],
\label{eq:s2_M_k21}
\\[1mm]
\kappa^{B_{\rm in}Q_{\rm in}}_{1\,2}
&=
\alpha\Big[
\alpha^2\,\kappa^{BQ}_{1\,2}
+\alpha\beta\,\kappa^{BQ}_{1\,1}
+2\alpha\beta\,\kappa(M^{BQ}_{1\,1},Q_{\rm tot})
+\beta(1-2\alpha)\,\kappa(M^{BQ}_{1\,2})
\Big],
\label{eq:s2_M_k12}
\\[1mm]
\kappa^{B_{\rm in}Q_{\rm in}}_{3\,1}
&=
\alpha\Big[
\alpha^3\,\kappa^{BQ}_{3\,1}
+3\alpha^2\beta\,\kappa^{BQ}_{2\,1}
+\alpha\beta(1-2\alpha)\,\kappa^{BQ}_{1\,1}
\nonumber\\
&\quad
+3\alpha^2\beta\,\kappa(M^{BQ}_{1\,1},B_{\rm tot},B_{\rm tot})
-\alpha\beta(6\alpha-3)\,\kappa(M^{BQ}_{2\,1},B_{\rm tot})
\nonumber\\
&\quad
+3\alpha\beta^2\,\kappa(M^{BQ}_{1\,1},B_{\rm tot})
+\beta(1-6\alpha\beta)\,\kappa(M^{BQ}_{3\,1})
\Big].
\label{eq:s2_M_k31}
\end{align}
Note that under charge idempotency the monomial-weighted multiplicities with repeated indices coincide, $M^{BQ}_{2\,1}=M^{BQ}_{1\,2}=M^{BQ}_{3\,1}=M^{BQ}_{1\,1}$; we keep the general labels to exhibit the block structure of Eq.~\eqref{eq:binomial-master}.
Since other terms in Eqs.~\eqref{eq:s2_M_k11}--\eqref{eq:s2_M_k31} have already been reduced using $b_{ip}^2=b_{ip}$ [e.g.\ $\kappa(M^{BQ}_{2\,0},Q_{\rm tot})\to\kappa^{BQ}_{1\,1}$ in Eq.~\eqref{eq:s2_M_k21}], these expressions do not apply verbatim to non-idempotent charges, for which one should return to the general formula~\eqref{eq:binomial-master}.
Substituting Eqs.~\eqref{eq:s2_M_k11}--\eqref{eq:s2_M_k31} into the SAM-3.0 mapping of Sec.~\ref{sec:multidim} yields the off-diagonal canonical cumulants. The leading result is Eq.~\eqref{eq:s2_ce_k11} of the main text, while at next order
\begin{equation}
\kappa^{B_{\rm in}Q_{\rm in}|BQ}_{2\,1}=
\alpha\beta(1-2\alpha)\,\mean{M^{BQ}_{1\,1}},
\label{eq:s2_ce_k21_app}
\end{equation}
which parallels the leading-order result~\eqref{eq:s2_ce_k11}, with the additional factor $(1-2\alpha)$ matching the Bernoulli skewness $c_3/c_2$.
Higher-order canonical cumulants follow analogously.
The cumulants $\kappa^{B_{\rm in}Q_{\rm in}|BQ}_{n\,1}$ remain proportional to $\mean{M^{BQ}_{1\,1}}$ at every order, $\kappa^{B_{\rm in}Q_{\rm in}|BQ}_{n\,1}=c_{n+1}(\alpha)\,\mean{M^{BQ}_{1\,1}}$, with the Bernoulli cumulants $c_n(\alpha)$ of Eq.~\eqref{eq:cn_examples}.
Joint cumulants of the charge-weighted multiplicities, which probe correlations among species carrying multiple conserved charges simultaneously, first enter at fourth order through the genuinely mixed cumulants such as $\kappa^{B_{\rm in}Q_{\rm in}|BQ}_{2\,2}$.

\bibliographystyle{JHEP}
\bibliography{main}

\end{document}